\documentclass[a4paper,11pt]{article}
\usepackage{jcappub} 
\usepackage{lineno}

\usepackage{amsmath}
\usepackage{bm}
\usepackage{url}
\usepackage{comment}

\title{{Impact of lensing bias on the cosmological dispersion measure}}
\author{Ryuichi Takahashi}
\affiliation{Faculty of Science and Technology, Hirosaki University, 3 Bunkyo-cho, Hirosaki, Aomori 036-8561, Japan}

\abstract{The cosmological dispersion measure (DM) as a function of redshift, derived from localized fast radio bursts (FRBs), has been used as a tool for constraining the cosmic ionized fraction and cosmological parameters.
For these purposes, the DM in a homogeneous cosmological model has typically been used, neglecting the inhomogeneity of matter distribution.
In this study, we derive a bias in the ensemble average of the DM over many FRBs owing to gravitational lensing by the inhomogeneous matter distribution based on cosmological perturbation theory.
We demonstrate that the ensemble average is $0.4 \, \%$--$1\,\%$ smaller than the DM in the corresponding homogeneous model for a source redshift of $z_{\rm s}=1$, according to recent cosmological hydrodynamic simulations of IllustrisTNG and BAHAMAS. 
This reduction occurs because light rays from FRBs tend to avoid high-density regions owing to lensing deflection. 
We also discuss another selection effect, magnification bias, where demagnified FRBs with low DMs, fainter than the detection threshold, are excluded from the observed sample, leading to a selective observation of magnified FRBs with high DMs.
Lensing bias, including magnification bias, must be considered to achieve percent level accuracy in the DM--redshift relation.}

\begin{document}
\maketitle
\flushbottom

\section{Introduction}

Fast radio burst (FRB) is a radio transient, with a typical duration of milliseconds, from a cosmological distance (e.g., recent reviews \cite{Bhandari2021,Xiao2021,Petroff2022,Zhang2023}).
Many theoretical models for FRBs have been proposed, but their physical origin is still elusive (e.g., \cite{Zhang2023}). 
From a frequency dependence of the arrival time, the dispersion measure (DM) is measured, which is the column density of free electrons along the light path to the source. 
If its host galaxy is identified, it is called a localized FRB, and its redshift can be determined. 
Currently, at least 40 localized FRBs have been identified (e.g., ~\cite{Keane2016,Chatterjee2017,Bhandari2022,Bhardwaj2023,Gordon2023,Ibik2024,Law2024}), with the highest redshift being $z_{\rm s}=1.0$~\cite{Ryder2023}.
The observed DM is influenced by contributions from the Milky Way (MW), the host galaxy, and cosmological objects such as the inter-galactic medium (IGM) and intervening galaxies.
Because the cosmological contribution is dominant for relatively high source redshifts ($z_{\rm s} \gtrsim 0.5$), we focus solely on the cosmological DM in this study.

The DM--redshift relation, measured from localized FRBs, has been utilized to constrain the ionized fraction in the IGM ($f_{\rm e}$)~\cite{Li2020,Lemos2023,Lin2023,Wang2023,Khrykin2024}, Hubble parameter ($H_0$)~\cite{Hagstotz2022,James2022,Wu2022,Fortunato2023,Liu2023,Wei2023,Gao2024}, and cosmic baryon abundance ($\Omega_{\rm b}$)~\cite{Keane2016,Macquart2020,Yang2022}. 
Because these parameters are completely degenerate (${\rm DM} \propto H_0 f_{\rm e} \Omega_{\rm b}$), external constraints or assumptions are required to determine one of them. 
When many localized FRBs are available at high redshifts ($z_{\rm s} \gtrsim 0.5$), the DM--$z$ relation can probe the properties of dark energy (e.g., \cite{Gao2014,Zhou2014,Walters2018}) and the history of cosmic reionization (e.g., \cite{Ioka2003,Inoue2004,Zheng2014,Caleb2019,Beniamini2021,Hashimoto2021}).
In all previous works, the DM was calculated using a homogeneous cosmological model.
In this study, we investigate the effects of matter inhomogeneity on the ensemble average of the DM over many FRBs.

The effects of inhomogeneous mass distribution on the luminosity distance $D_{\rm L}$ have been extensively studied for supernova (SN) cosmology, utilizing second-order cosmological perturbation theory (e.g., \cite{Adamek2019,Barausse2005,Ben-Dayan2012,Ben-Dayan2013JCAP,Ben-Dayan2013PRL,Bonvin2015b,Brenton2021,Fleury2017,KP2016,Marozzi2015, Umeh2014,Umeh2014II}; also the recent review \cite{Helbig2020}).
Matter inhomogeneity causes a scatter and a bias in $D_{\rm L}$. 
The bias arises from the second-order term of density fluctuations because the first-order term averages out over many sources.
For high redshifts ($z_{\rm s} \gtrsim 0.3$), the dominant contribution is gravitational lensing, while for low redshifts ($z_{\rm s} \lesssim 0.3$), it is the peculiar motions of the observer and the source \cite{Ben-Dayan2013JCAP,Ben-Dayan2013PRL}.
In this study, we focus on the lensing contribution, with other second-order effects briefly discussed in Subsection \ref{subsec:discussion1}.
The ensemble average of $D_{\rm L}$ in an inhomogeneous universe, $\langle D_{\rm L} \rangle$, and that in the corresponding homogeneous model, $\bar{D}_{\rm L}$, follow a simple relation:
 $\langle D_{\rm L} \rangle = [ \, 1+ ({3}/{2}) \langle \kappa^2 \rangle ] \bar{D}_{\rm L}$, where $\kappa$ is the convergence field (e.g., \cite{Clarkson2014,KP2016}).
Given that the variance of $\kappa$ is estimated as $\langle \kappa^2 \rangle \approx (2$--$8) \times 10^{-3} \, z_{\rm s}^2$~\cite{Holz2005,Jonsson2010,Shah2022,Shah2024}, the lensing bias increases $D_{\rm L}$ by ${\mathcal O}(0.1 \%$--$1 \%)$ for $z_{\rm s} \gtrsim 0.4$.
This bias will need to be accounted for in future SN cosmology analyses (e.g., \cite{Adamek2019,Shah2023}).

Kaiser \& Peacock (2016) \cite{KP2016} (hereafter KP16) evaluated the lensing bias in $\langle D_{\rm L} \rangle$ using standard perturbation theory.
By solving the null-geodesic equation, they demonstrated that light rays propagate through underdense regions on average because they tend to avoid matter clumps owing to lensing deflection. 
They also analytically derived that the average convergence $\kappa$ is negative, satisfying the simple relation $\langle \kappa \rangle = -2 \langle \kappa^2 \rangle$.
This originates from flux conservation\footnote{From flux conservation, the average lensing magnification is unity, $\langle \mu \rangle =1$. Using $\mu=1/[(1-\kappa)^2-\gamma^2]$ ($\gamma$ is the shear field, satisfying $\langle \gamma^2 \rangle = \langle \kappa^2 \rangle$) with $|\kappa|,|\gamma| \ll 1$, one obtains $\langle \kappa \rangle = -2 \langle \kappa^2 \rangle$. Similarly, the bias in $D_{\rm L}$ is obtained as $\langle D_{\rm L} \rangle = \langle \mu^{-1/2} \rangle \bar{D}_{\rm L} \simeq [1+(3/2) \langle \kappa^2 \rangle] \bar{D}_{\rm L}$.} \cite{Weinberg1976}.
This relation is also confirmed by cosmological ray-tracing simulations \cite{RT2011} and for weak lensing of gravitational waves in wave optics (i.e., without the geometrical-optics approximation)~\cite{Mizuno2023,Mizuno2024}.

In this study, we calculate the lensing bias in the average of the DM, $\langle {\rm DM} \rangle$, in an inhomogeneous universe where the total matter (dark matter and baryons) and free-electron distributions are inhomogeneous.
A standard deviation of the DM caused by matter inhomogeneity has been previously studied (e.g., \cite{McQuinn2014,Dolag2015,Shirasaki2017,Jaroszynski2019,TI2021}), but the bias has not yet been examined.
We use the standard cosmological perturbation theory following the KP16 procedure. 
Based on previous studies on $\langle D_{\rm L} \rangle$, we expect $\langle {\rm DM} \rangle$ to be smaller than the prediction from the homogeneous model.
Section \ref{sec:2} details our approach, where we integrate the free-electron density along null geodesics in curved spacetime to compute the DM and then expand it in terms of density fluctuations. 
We derive a leading correction term in $\langle {\rm DM} \rangle$ determined by the cross-power spectrum of matter and free electrons (Appendix \ref{sec:appendix1}), utilizing data from recent hydrodynamic simulations of IllustrisTNG\footnote{\url{https://www.tng-project.org/}} and BAHAMAS\footnote{\url{https://www.astro.ljmu.ac.uk/~igm/BAHAMAS/}}.
Section \ref{sec:3} presents numerical results for $\langle {\rm DM} \rangle$, comparing it with predictions from the homogeneous model. 
We also discuss the influence of baryonic feedback on $\langle {\rm DM} \rangle$.
Section \ref{sec:mag_bias} addresses another selection effect, magnification bias, and its impact on $\langle {\rm DM} \rangle$.
Section \ref{sec:discussion} discusses other second-order effects aside from lensing bias (Subsection \ref{subsec:discussion1}) and provides insights into lensing bias in rotation measures (Subsection \ref{subsec:discussion2}).
Finally, Section \ref{sec:conclusion} summarizes the findings of this study.

In this paper, for any arbitrary quantity $f$, $\langle f \rangle$ denotes its ensemble average in an inhomogeneous universe, while $\overline{f}$ represents its value in the corresponding homogeneous model. 
We adopt a spatially flat $\Lambda$CDM model consistent with the {\it Planck} 2015 best-fitting parameters~\citep{Planck2016}: matter density $\Omega_{\rm m}=1-\Omega_\Lambda=0.3089$, baryon density $\Omega_{\rm b}=0.0486$, Hubble parameter $h=0.6774$, spectral index $n_{\rm s}=0.9667$, and amplitude of matter density fluctuations on the scale of $8 \, h^{-1}$Mpc $\sigma_8=0.8159$.
This model is identical to the one used in IllustrisTNG.
All physical quantities such as length, wavenumber, and number density are expressed in comoving units.

\section{Ensemble average of the DM}
\label{sec:2}

\subsection{Light ray path in an inhomogeneous universe}

\begin{figure}
  \centering
  \includegraphics[width=7cm, clip]{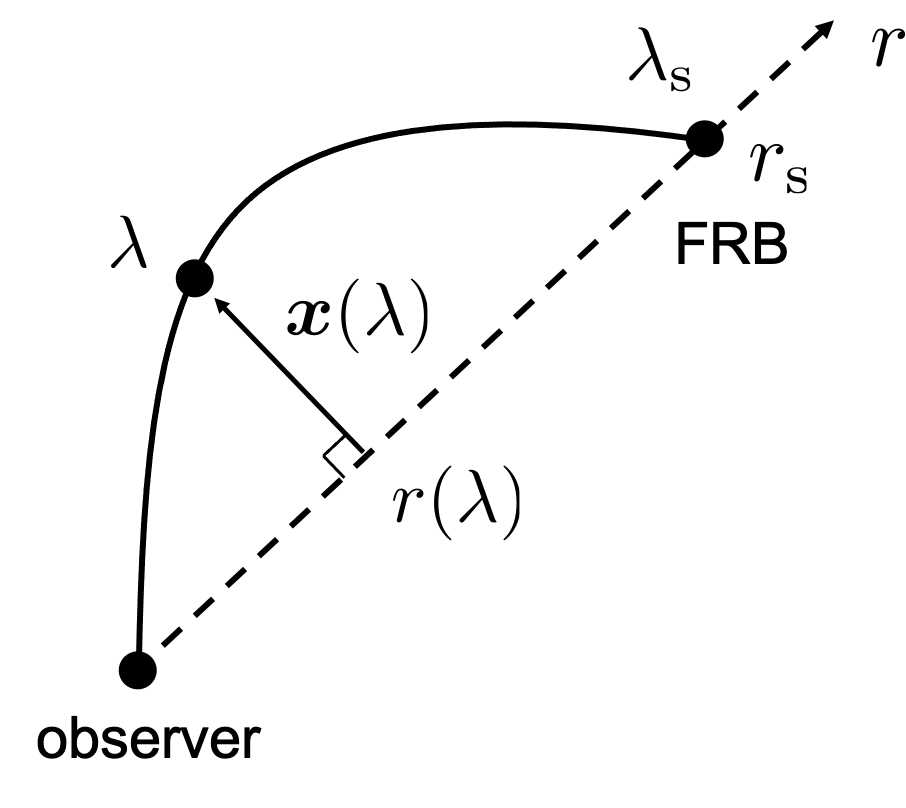}  
  \caption{
    The solid curve represents the path of a light ray toward the FRB, parameterized by the affine parameter (path length) $\lambda$: $\lambda=0$ at the observer and $\lambda=\lambda_{\rm s}$ at the source. The dashed line represents the radial coordinate $r$ toward the source, and $\bm{x}$ denotes a two-dimensional coordinate perpendicular to the $r$-axis.}
  \label{fig_config}
\end{figure}

This subsection briefly explores the path of a light ray in an inhomogeneous universe (e.g., Appendices of KP16; a review \cite{BS2001}; a textbook \cite{Dodelson2020}). 
Figure \ref{fig_config} illustrates the coordinate system used. 
The $r$ axis represents the radial coordinate (comoving distance) connecting the observer and the FRB.
The coordinate $r$ is expressed as a function of redshift $z$: $r(z)=c \int_0^z {\rm d}z^\prime/H(z^\prime)$, where $H(z)$ is the Hubble expansion rate.
The distance to the source is $r_{\rm s}=r(z_{\rm s})$.
The two-dimensional coordinate $\bm{x}$ is perpendicular to the $r$-axis.
The spatial coordinate is denoted by $\bm{r}=(r,\bm{x})$.

In a homogeneous universe, a light ray from the source travels along the $r$-axis.
In an inhomogeneous scenario, the propagation of a light ray is affected by gravitational deflections caused by matter clumps.
The matter inhomogeneity generates a gravitational potential $\phi=\phi(\bm{r};z)$, assuming weak gravity ($|\phi|/c^2 \ll 1$).
We adopt the Newtonian gauge for the metric perturbation and neglect the peculiar velocities of the observer and source.
The path of a light ray is determined by the geodesic equation with the affine parameter $\lambda$: ${\rm d}^2 \bm{r}/{\rm d} \lambda^2 = -(2/c^2) \, \nabla_\perp \phi$ where $\nabla_\perp$ is derivative in the direction perpendicular to the path.
Here, $\lambda$ is the path length from the observer.
The geodesic equation is solved under the boundary condition that the positions at both endpoints are fixed to those of the observer and the source~(Appendix E of KP16):
\begin{align}
        \frac{{\rm d} \bm{x}}{{\rm d}\lambda}(\lambda) &= -\frac{2}{c^2} \int_0^{\lambda} \! {\rm d}\lambda^\prime \nabla_{\bm{x}^\prime} \phi(\bm{r}^\prime;z^\prime) 
    + \frac{2}{c^2} \frac{1}{\lambda_{\rm s}} \int_0^{\lambda_{\rm s}} \! {\rm d}\lambda^\prime (\lambda_{\rm s}-\lambda^\prime) \nabla_{\bm{x}^\prime} \phi(\bm{r}^\prime; z^\prime),  \nonumber \\
    \bm{x}(\lambda) &= -\frac{2}{c^2} \int_0^{\lambda} \! {\rm d}\lambda^\prime (\lambda-\lambda^\prime) \nabla_{\bm{x}^\prime} \phi(\bm{r}^\prime;z^\prime) 
    + \frac{2}{c^2} \frac{\lambda}{\lambda_{\rm s}} \int_0^{\lambda_{\rm s}} \! {\rm d}\lambda^\prime (\lambda_{\rm s}-\lambda^\prime) \nabla_{\bm{x}^\prime} \phi(\bm{r}^\prime; z^\prime), 
\label{eq:geodesic}
\end{align}
where $\bm{r}^\prime=\bm{r}(\lambda^\prime)=(r(\lambda^\prime),\bm{x}(\lambda^\prime))$, $z^\prime=z(\lambda^\prime)$, and $\nabla_{\bm{x}}$ ($\simeq \nabla_\perp$) is the gradient with respect to $\bm{x}$.
Equation (\ref{eq:geodesic}) includes terms up to the first order of $\phi/c^2$.
Because the gravitational potential is weak, we obtain $|\bm{x}| \ll r$. 

A light ray can also experience deflection by an inhomogeneous plasma density caused by spatial variations in the refractive index (plasma lensing).
We have disregarded this effect because the deflection angle is typically negligible, except in cases involving high-density clumps at low radio frequencies $\nu$.   
For lensing by an intervening galaxy, the deflection angle is typically $\approx 1 \, {\rm arcsec}$ for gravitational lensing and $\approx 10^{-3} \, (\nu/400 {\rm MHz})^{-2} \, {\rm arcsec}$ for plasma lensing~(e.g., \cite{Er2014}).

\subsection{The DM in an inhomogeneous universe}

The DM represents the column density of free electrons along the path to the source\footnote{The DM--$z$ relation is sometimes referred to as the Macquart relation~\cite{Macquart2020}, although it had been 
developed earlier in previous studies (e.g., \cite{HS1965,Ginzburg1973,Palmer1993,Ioka2003,Inoue2004,DZ2014}).} (e.g., \cite{Palmer1993,Ioka2003,Inoue2004}):
\begin{equation}
   {\rm DM}(z_{\rm s}) = \int_0^{\lambda_{\rm s}} \! {\rm d}\lambda \, n_{\rm e}(\bm{r}; z) \, (1+z).
\label{eq:DM}
\end{equation}
The integration is conducted along the geodesic (\ref{eq:geodesic}).
The free-electron density $n_{\rm e}$ is separated into its spatial mean $\bar{n}_{\rm e}$ and fluctuations $\delta_{\rm e}$:
\begin{equation}
   n_{\rm e}(\bm{r};z) = \bar{n}_{\rm e}(z) \left[ 1 + \delta_{\rm e}(\bm{r};z) \right].
\label{eq:ne}
\end{equation}
The spatial average of the second term vanishes, $\langle \delta_{\rm e} \rangle=0$.
The mean free-electron number density is expressed as~(e.g., \cite{DZ2014}):
\begin{equation}
    \bar{n}_{\rm e}(z) = \frac{3 H_0^2}{8 \pi G} \frac{\Omega_{\rm b}}{m_{\rm p}} \left( X(z) + \frac{1}{2} Y(z) \right) f_{\rm e}(z),
\label{eq:ne_mean}
\end{equation}
where $m_{\rm p}$ is the proton mass, $X$ and $Y$ are the mass fractions of hydrogen and helium ($X \simeq 0.75$ and $Y \simeq 0.25$), and $f_{\rm e}$ is the ionized fraction ($f_{\rm e} \approx 0.7$--$0.9$).
In the homogeneous model, the DM simplifies to
\begin{align}
 \overline{\rm DM}(z_{\rm s}) &= \int_0^{r_{\rm s}} \! {\rm d}r \, \bar{n}_{\rm e}(z) \, (1+z),  \nonumber  \\
 &= \int_0^{z_{\rm s}} \! \frac{c \, {\rm d}z}{H(z)} \, \bar{n}_{\rm e}(z) \, (1+z).
\label{eq:homoDM}
\end{align}
The integration is done along the $r$-axis.

Next, we derive a leading correction term in the DM arising from second-order perturbations ($\phi/c^2$ and $\delta_{\rm e}$).
In Eq.~(\ref{eq:DM}), because $|\bm{x}| \ll r$, $n_{\rm e}$ can be expanded up to the linear order of $\bm{x}$:
\begin{align}
    n_{\rm e}(\bm{r};z) &\simeq \left. n_{\rm e}(\bm{r};z)\right|_{\bm{x}=0} + \left. \bm{x}\cdot\nabla_{\bm{x}} n_{\rm e}(\bm{r};z)\right|_{\bm{x}=0}, \nonumber \\
    &= \bar{n}_{\rm e}(z) \left[ \, 1+ \left. \delta_{\rm e}(\bm{r};z)\right|_{\bm{x}=0} + \left. \bm{x}\cdot\nabla_{\bm{x}} \delta_{\rm e}(\bm{r};z)\right|_{\bm{x}=0} \right]. 
\label{eq:expand_ne}
\end{align}
The integral along the geodesic in Eq.~(\ref{eq:DM}) is replaced by integration along the $r$-axis:
\begin{align}
     {\rm d}\lambda &= \sqrt{\, 1+ \left. \left( \frac{{\rm d}\bm{x}}{{\rm d}\lambda} \right)^2\right|_{\bm{x}=0}} ~{\rm d}r, \nonumber \\
     & \simeq {\rm d}r + \frac{1}{2} \left. \left( \frac{{\rm d}\bm{x}}{{\rm d}\lambda} \right)^2 \right|_{\bm{x}=0} {\rm d}r.
\label{eq:expand_lambda}
\end{align}
Using Eqs.~(\ref{eq:DM}), (\ref{eq:expand_ne}) and (\ref{eq:expand_lambda}), we derive the ensemble average of the DM as
\begin{align}
    \langle {\rm DM} \rangle  (z_{\rm s}) &= \int_0^{r_{\rm s}} \! {\rm d}r \, \bar{n}_{\rm e}(z) \, (1+z) 
    + \int_0^{r_{\rm s}} \! {\rm d}r \, \bar{n}_{\rm e}(z) \, (1+z) \, \left\langle \left. \bm{x}\cdot\nabla_{\bm{x}} n_{\rm e}(\bm{r};z)\right|_{\bm{x}=0}  \right\rangle    \nonumber \\
    &~~+ \frac{1}{2} \int_0^{r_{\rm s}} \! {\rm d}r \, \bar{n}_{\rm e}(z) \, (1+z) \left\langle \left. \left( \frac{{\rm d}\bm{x}}{{\rm d}\lambda} \right)^2 \right|_{\bm{x}=0} \right\rangle.
\label{eq:expand_DM}
\end{align}
The first term corresponds to the DM in the homogeneous model (\ref{eq:homoDM}). 
The second and third terms represent the leading correction terms. 
Appendix \ref{sec:appendix1} provides a detailed calculation of these terms; the second term is negative\footnote{The vector $\bm{x}$ points from high to low density, whereas the direction of $\nabla_{\bm{x}} n_{\rm e}$ is the opposite; thus, the second term is negative.}, whereas the third is positive.
Because the third term is $\approx 10^{-7}$ ($10^{-5}$) times smaller than the first (second) term for $z_{\rm s}=0$--$3$, we disregard the third term in subsequent analysis.

In summary, the ensemble average of the DM is written as
\begin{equation}
    \langle {\rm DM} \rangle (z_{\rm s}) = \overline{\rm DM}(z_{\rm s}) + \Delta {\rm DM}(z_{\rm s}),
\label{eq:ensemble_DM}
\end{equation}
with 
\begin{align}
    \Delta {\rm DM}(z_{\rm s}) &= -\frac{3H_0^2 \Omega_{\rm m}}{2 \pi c^2} \int_0^{z_{\rm s}} \!\! \frac{c \, {\rm d}z}{H(z)}  \, \bar{n}_{\rm e}(z) \, (1+z)^2 \, \frac{r(z) \left( r_{\rm s}-r(z) \right)}{r_{\rm s}} 
     \! \int_0^\infty \! {\rm d}k_\perp k_\perp P_{\rm me}(k_\perp;z),
\label{eq:dDM}
\end{align}
where $P_{\rm me}$ is the cross-power spectrum of matter and free electrons (defined in Eq.~(\ref{eq:cross_pk})); therefore, $\Delta {\rm DM}$ originates from the cross-correlation between these two fields.
The distributions of matter and free electrons determine $\bm{x}$ and $\nabla_{\bm{x}} n_{\rm e}$, respectively, in Eq.~(\ref{eq:expand_DM}). Our result (\ref{eq:ensemble_DM}) is consistent with the previous one for the average convergence, $\langle \kappa \rangle = -2 \langle \kappa^2 \rangle$, when $\kappa$ is used instead of the DM (see also Appendix \ref{sec:appendix1}).
According to Eq.~(\ref{eq:dDM}), the density fluctuations approximately halfway to the source (i.e., $r(z) \approx r_{\rm s}/2$) contribute the most to $\Delta {\rm DM}$.

\subsection{The cross-power spectrum of matter and free electrons}

\begin{figure}
  \centering
  \includegraphics[width=15.5cm, clip]{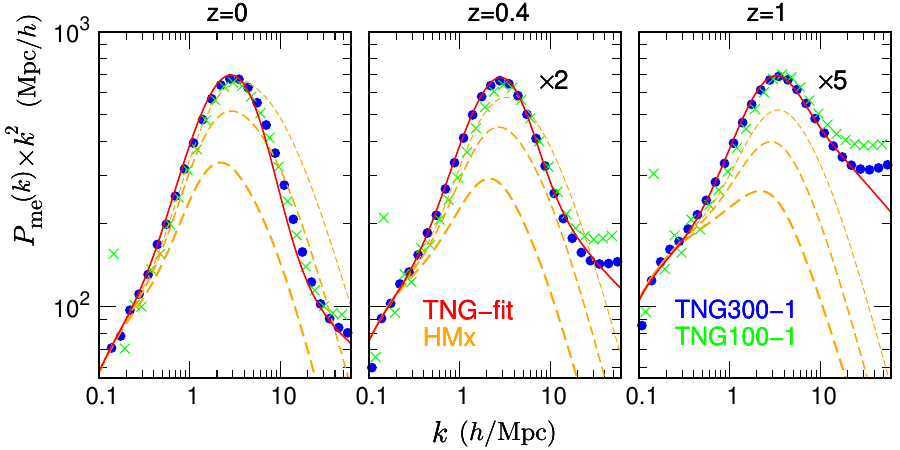}  
  \caption{
    Cross-power spectrum of matter and free electrons at $z=0$, $0.4$, and $1$. The blue and green symbols represent results from hydrodynamic simulations TNG300-1 and TNG100-1, respectively. The red curves depict our theoretical model based on the free-electron bias (\ref{eq:pk_theory}) fitted to the TNG results. The dashed orange curves show predictions from \textsc{HMx}~\cite{Mead2020}, calibrated with BAHAMAS. The three orange curves differ in the AGN feedback temperature: $\log_{10} (T_{\rm AGN}/{\rm K})=7.6$ (thin), $7.8$ (average), and $8.0$ (thick), where the higher temperature corresponds to the stronger feedback. In the middle and right panels, the amplitudes are multiplied by $2$ and $5$, respectively, to make the peak heights uniform. 
    }
  \label{fig_pk}
\end{figure}

This subsection discusses two theoretical models of $P_{\rm me}$
in Eq.~(\ref{eq:dDM}).  
The first one is our fitting function of $P_{\rm me}$ based on the hydrodynamic simulations of IllustrisTNG~\citep{Marinancci2018,Naiman2018,Nelson2018,Phillepich2018,Springel2018,Nelson2019}. 
The other is a numerical scheme of \textsc{HMx}\footnote{\url{https://github.com/tilmantroester/pyhmcode}}~\cite{Mead2020,Troster2022} calibrated with another hydrodynamic simulation suite of BAHAMAS~\citep{McCarthy2017,McMarthy2018}.
A comparison of these different models helps to understand the baryonic feedback effect on $P_{\rm me}$.

We measure $P_{\rm me}$ in the public IllustrisTNG data set~\cite{Nelson2019} to make its fitting function, following the same procedure as in our previous work~\cite{TI2021} (their Subsection 3.3). 
The simulations follow the gravitational clustering of matter (dark matter and baryons), as well as astrophysical processes such as star and galaxy formation, and stellar and active
galactic nucleus (AGN) feedback. 
The gravitational evolution and magneto-hydrodynamic processes were computed with the moving-mesh code \textsc{AREPO}~\cite{Springel2010}. 
We analyzed two simulations with different box sizes: TNG300-1 and TNG100-1. 
TNG300-1 (TNG100-1) comprises $2500^3$ ($1820^3$) dark matter particles and a nearly equivalent number of baryon particles in a cubic box with a side length of 205 (75) $h^{-1} {\rm Mpc}$; thus, TNG100-1 offers higher spatial and mass resolutions.
The baryon particles consist of gas, stars, and black holes, with free electrons included in the gas particles.
To calculate the matter density contrast $\delta_{\rm m}$, we assign the masses of dark matter and baryon particles to regular grid cells using the cloud-in-cell interpolation with the interlacing scheme (e.g., \cite{HE1981,Jing2005,Sefusatti2016}).
Similarly, we calculate the free-electron density contrast $\delta_{\rm e}$ by assigning the masses of free electrons to the grid cells.
Fourier transforms of these density fields, $\tilde{\delta}_{\rm m}(\bm{k})$ and $\tilde{\delta}_{\rm e}(\bm{k})$, are computed using the fast Fourier transform\footnote{{FFTW} (Fastest Fourier Transform in the West) at \url{http://www.fftw.org}.}. 
The power spectrum is then measured as
\begin{equation}
    P_{\rm me}(k;z) = \frac{1}{N_{\rm mode}} \sum_{|\bm{k}^\prime| \in k} {\rm Re}\left[  \tilde{\delta}_{\rm m}(\bm{k}^\prime) \tilde{\delta}^*_{\rm e}(\bm{k}^\prime) \right],
\label{PS}
\end{equation}
where the summation is performed in the spherical shell $k-\Delta k/2<|\bm{k}^\prime|<k+\Delta k/2$ with a bin width of $\Delta \log_{10} k=0.1$, and $N_{\rm mode}$ represents the number of Fourier modes in the shell.
The power spectrum is reliable up to the particle Nyquist wavenumber $k_{\rm Nyq}$, determined by the mean separation of gas particles $r_{\rm gas}$: $k_{\rm Nyq}=\pi/r_{\rm gas}= 38.3 \, (76.2) \, h \, {\rm Mpc}^{-1}$ for TNG300-1 (TNG100-1). 
We measured $P_{\rm me}$ at $z=0, 0.2, 0.4, 0.7, 1, 2$ and $3$.

The TNG team also conducted dark matter-only (DMO) runs, where the box sizes and the number of dark matter particles matched those in TNG300-1 and TNG100-1, respectively. 
These runs include N-body particles representing dark matter and baryons, focusing solely on gravitational evolution.
We measure the matter power spectra in the DMO runs, denoted as $P_{\rm DMO}$, using the identical procedure as for $P_{\rm me}$.
Subsequently, we define the free-electron bias: 
\begin{equation}
  b_{\rm e}(k;z) \equiv \frac{P_{\rm me}(k;z)}{P_{\rm DMO}(k;z)}.
\label{eq:pk_theory}
\end{equation}
The fitting function of $b_{\rm e}$ is detailed in Appendix \ref{sec:be}.
It is worth noting that the free-electron bias (\ref{eq:pk_theory}) differs slightly from our previous definition~\cite{TI2021}, i.e., $b_{\rm e}^{\rm (prev)}(k;z) \equiv [P_{\rm e}(k;z)/P_{\rm DMO}(k;z)]^{1/2}$, where $P_{\rm e}(k;z)$ is the auto-power spectrum of free electrons. 
The discrepancy is noticeable only on small scales ($k \gtrsim 1 h {\rm Mpc}^{-1}$), as discussed in Appendix \ref{sec:be}.
 
Moving on to \textsc{HMx}, the BAHAMAS simulations were conducted using the smoothed-particle hydrodynamic code \textsc{Gadget}-3~\cite{Springel2005}, incorporating subgrid models to simulate baryonic processes. 
The stellar and AGN feedback models were calibrated to match observations of the galaxy's stellar mass function and the hot gas fraction in galaxy groups and clusters.
AGN feedback strength is specified by the heating temperature: $\log_{10} (T_{\rm AGN}/{\rm K})=7.6, 7.8$, and $8.0$ where $7.8$ is their fiducial value consistent with the observed hot gas fraction of groups and clusters~\cite{McCarthy2017}. 
There are $1024^3$ dark matter particles and an equivalent number of baryon particles in a box volume of $(400 \, h^{-1} {\rm Mpc})^3$.
\textsc{HMx} utilizes the halo model framework (e.g., \cite{CS2002,Arico2020,Shirasaki2022,Asgari2023}), calibrated to reproduce auto- and cross-power spectra of five fields: total matter, cold dark matter, gas, stars, and electron pressure, measured in BAHAMAS.
The calibration range is $k=0.015$--$7 \, h {\rm Mpc}^{-1}$ and $z=0$--$1$.
We derive $P_{\rm me}$ from the cross-power spectrum of matter and gas in \textsc{HMx}, denoted as $P_{\rm mg}^{\rm HMx}$, assuming that free electrons exactly trace gas (i.e., $\delta_{\rm e}=\delta n_{\rm e}/\bar{n}_{\rm e}$ $=\delta \rho_{\rm gas}/\bar{\rho}_{\rm gas}$).
Because the normalization of $P_{\rm mg}^{\rm HMx}$ differs from our $P_{\rm me}$, we rescale it\footnote{These two power spectra are related via $P_{\rm me}(k;z)=(\bar{\rho}_{\rm m}/\bar{\rho}_{\rm gas}(z)) \, P_{\rm mg}^{\rm HMx}(k;z)$ $=(\Omega_{\rm m}/\Omega_{\rm b}) \, [1-(\Omega_{\rm m}/\Omega_{\rm b})(\bar{\rho}_{\rm star}(z)/\bar{\rho}_{\rm m})]^{-1} P_{\rm mg}^{\rm HMx}(k;z)$, where $\bar{\rho}_{\rm m}, \bar{\rho}_{\rm gas}$ and $\bar{\rho}_{\rm star}$ are the mean densities of matter, gas, and stars, respectively ($\bar{\rho}_{\rm m}$ is constant while the others are functions of $z$). The ratio $\bar{\rho}_{\rm star}(z)/\bar{\rho}_{\rm m}$ is obtained by averaging the stellar fraction (Eq. (27) in \cite{Mead2020}) over all halo masses.}.
Previously, the \textsc{HMx} power spectrum was used to calculate the covariance of the DM~\cite{Reischke2023} and the angular cross-power spectrum of the DM and $\kappa$~\cite{Reischke2023b}.

Figure \ref{fig_pk} presents $P_{\rm me}$ at $z=0, 0.4$, and $1$.
The blue and green symbols denote TNG measurements. 
The TNG300-1 and TNG100-1 show consistent results for $k \lesssim 10 \, h {\rm Mpc}^{-1}$. 
However, the TNG100-1 gives slightly higher amplitudes for $k \gtrsim 10 \, h {\rm Mpc}^{-1}$, probably owing to the finer spatial resolution.
The red curves are our theoretical model, $P_{\rm me}(k;z) = b_{\rm e}(k;z) P_{\rm DMO}(k;z)$, using the fitting function of $b_{\rm e}$ and  \textsc{Halofit}~\citep{Smith2003,RT2012} for the non-linear matter power spectrum $P_{\rm DMO}$.
The dashed orange curves are the \textsc{HMx} results.
All TNG and \textsc{HMx} results agree at the largest scale ($k<1 \, h {\rm Mpc}^{-1}$) because the free electrons trace the underlying dark matter irrespective of baryonic physics~\cite{TI2021}.
The free-electron bias is almost unity at such a large scale (Appendix \ref{sec:be}).
For $k > 1 \, h {\rm Mpc}^{-1}$, $P_{\rm me}$ shows suppression influenced by AGN feedback strength; higher $T_{\rm AGN}$ leads to stronger suppression, consistent with trends seen in the matter power spectrum (e.g., \cite{Chisari2019}). 
From the peak of $k^2 P_{\rm me}(k)$ in Fig.~\ref{fig_pk}, the density fluctuations of $k=1$--$10 \, h {\rm Mpc}^{-1}$ contribute the most to $\Delta {\rm DM}$ in Eq.~(\ref{eq:dDM}).

Let us briefly summarize the current observational constraints on the $T_{\rm AGN}$ parameter.
Chen et al. (2023)~\cite{Chen2023} detected baryon feedback suppression on the matter power spectrum using cosmic shear measurements from the Dark Energy Survey (DES) year 3 data.
They found that the suppression was consistent with values of $\log_{10}(T_{\rm AGN}/{\rm K})=7.8$ and $8.0$ (their Fig.~8), which is consistent with other cosmic shear analyses~\cite{Arico2023,Xu2023}.
On the other hand, Terasawa et al. (2024)~\cite{Terasawa2024} performed a similar analysis using the Subaru Hyper Suprime-Cam three-year data, probing down to very small scales ($0.28$ arcmin). 
Their findings suggest a preference for weaker feedback with $\log_{10}(T_{\rm AGN}/{\rm K})=7.6$ (their Fig. 11; also discussed in \cite{Garcia2024}).
Tr{\"o}ster et al. (2022)~\cite{Troster2022} performed a combined analysis of cosmic shear from the Kilo-Degree Survey and the thermal Sunyaev--Zel'dovich data from Planck and the Atacama Cosmology Telescope.
Their constraint is $\log_{10}(T_{\rm AGN}/{\rm K})=7.96^{+0.20}_{-0.48}$. 
Ferreira et al. (2024)~\cite{Ferreira2023} measured a cross-power spectrum of weak lensing in DES Y3 and the diffuse X-ray background in the Roentgensatellit (ROSAT), providing a very precise constraint of $\log_{10}(T_{\rm AGN}/{\rm K})= 7.998 \pm 0.008$. 
In summary, the constraints do not converge well among the different measurements.

\section{Results}
\label{sec:3}

\begin{figure}
  \centering
  \includegraphics[width=10cm, clip]{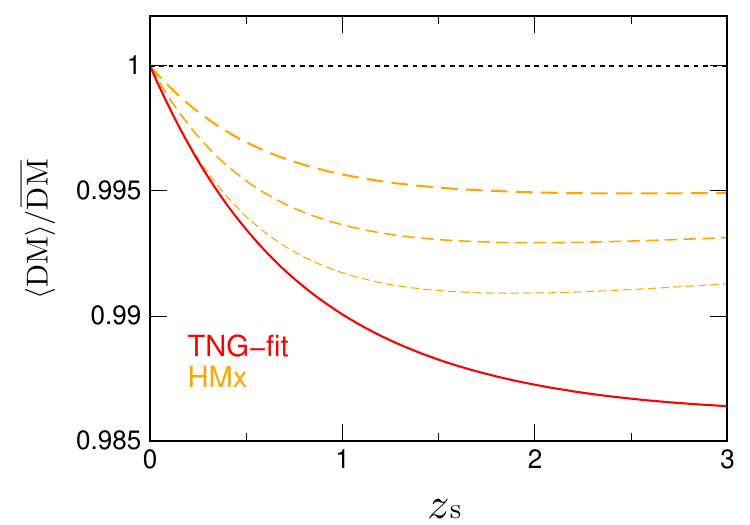} 
  \caption{
    The ensemble average of the DM in an inhomogeneous universe $\langle {\rm DM} \rangle$ normalized by that in the corresponding homogeneous model $\overline{\rm DM}$ as a function of source redshift $z_{\rm s}$. These curves are distinguished by the $P_{\rm me}$ model: the solid red curve uses the fitting function (\ref{eq:pk_theory}) for TNG, and the dashed orange curves use \textsc{HMx} with $\log_{10}(T_{\rm AGN}/{\rm K})=7.6$ (thin), $7.8$ (average) and $8.0$ (thick). 
    The dashed curves for $z_{\rm s}>1$ are obtained from an extrapolation of \textsc{HMx} (which is calibrated up to $z=1$). 
    }
  \label{fig_averageDM}
\end{figure}

This section presents our main results for $\langle {\rm DM} \rangle$ as described in Eq.~(\ref{eq:ensemble_DM}).
Figure \ref{fig_averageDM} illustrates $\langle {\rm DM} \rangle$ normalized by $\overline{\rm DM}$ as a function of $z_{\rm s}$.
Here, we set $f_{\rm e}=0.83$ and $X=1-Y=0.76$ in Eq.~(\ref{eq:ne_mean}), noting that the ratio $\langle {\rm DM}  \rangle/\overline{\rm DM}$ is independent of $f_{\rm e}, X$ and $Y$ as long as these quantities are constant.
The TNG results exhibit a stronger lensing bias owing to weaker baryon feedback.
For TNG, $\langle {\rm DM} \rangle$ is $1 \, \%$ ($0.6 \, \%$) smaller than $\overline{\rm DM}$ at $z_{\rm s}=1$ ($z_{\rm s}=0.5$), with the bias becoming more pronounced at higher $z_{\rm s}$.
For \textsc{HMx}, $\langle {\rm DM} \rangle$ is $0.4 \, \%$--$0.8 \, \%$ smaller than $\overline{\rm DM}$ at $z_{\rm s}=1$, depending on the value of $T_{\rm AGN}$. 
Notably, \textsc{HMx} is calibrated only up to $z=1$; thus, the results for $z_{\rm s}>1$ are extrapolated.
If the baryonic feedback is ignored (i.e., $b_{\rm e}=1$), $\Delta {\rm DM}$ becomes approximately twice the TNG result.
In summary, the conventional formula for $\overline{\rm DM}$ slightly overestimates the DM owing to the gravitational lensing effect, where light rays tend to avoid dense matter clumps.  

We estimate the number of localized FRBs needed to detect the lensing bias in the DM--$z$ relation for $z_{\rm s}=0.5$ and $1$. 
While the total DM comprises contributions from the MW, host galaxy, and cosmological sources, the lensing bias impacts the cosmological component. 
Referring to Fig.~\ref{fig_averageDM}, the discrepancy between $\langle {\rm DM} \rangle$ and $\overline{\rm DM}$ is 
\begin{align}
    \langle {\rm DM} \rangle(z_{\rm s})
     - \overline{\rm DM}(z_{\rm s}) &\approx -3.0 \, {\rm pc/cm}^3 \left( \frac{\langle {\rm DM} \rangle / \overline{\rm DM}}{0.994} \right)  ~~{\rm for} ~z_{\rm s}=0.5, \nonumber \\
    &\approx -10 \, {\rm pc/cm}^3 \left( \frac{\langle {\rm DM} \rangle / \overline{\rm DM}}{0.99} \right)  ~~{\rm for} ~z_{\rm s}=1,
\label{eq:diff_DM}
\end{align}
for the TNG result.
Here, we have used the approximate relation $\overline{\rm DM}(z_{\rm s}) \approx 1000 \, z_{\rm s}$ ${\rm pc/cm}^3$ (e.g., \cite{Ioka2003}).
The standard deviation of the measured DM ($\sigma_{\rm DM}$) arises from cosmological contribution ($\sigma_{\rm DM,cosmo}$) and host-galaxy contribution ($\sigma_{\rm DM,host}$): $\sigma_{\rm DM}=(\sigma_{\rm DM,cosmo}^2+\sigma_{\rm DM,host}^2)^{1/2}$.
Using $\sigma_{\rm DM,cosmo} \approx 200 \, z_{\rm s}^{1/2}$ ${\rm pc/cm}^3$ (e.g., \cite{McQuinn2014,Kumar2019,TI2021}) and $\sigma_{\rm DM,host} \approx 100 \, (1+z_{\rm s})^{-1}$ ${\rm pc/cm}^3$ (e.g., \cite{Macquart2020,Mo2023,Theis2024}), the standard deviation is
\begin{align}
    \sigma_{\rm DM}(z_{\rm s}) &\approx 160 \, {\rm pc/cm}^3 ~~{\rm for} ~z_{\rm s}=0.5, \nonumber \\
    &\approx 210 \, {\rm pc/cm}^3 ~~{\rm for} ~z_{\rm s}=1.
\label{eq:sigma_DM}
\end{align}
Using many FRBs, the standard deviation decreases inversely proportional to the square root of their count. 
Therefore, according to Eqs.~(\ref{eq:diff_DM}) and (\ref{eq:sigma_DM}), when approximately $3000$ and $400$ localized FRBs are available for $z_{\rm s}=0.5$ and $1$, respectively, the lensing bias will affect $\langle {\rm DM} \rangle$ at the $1 \sigma$ error level. 
In the near future, instruments such as the Canadian Hydrogen Intensity Mapping Experiment FRB instrument (CHIME/FRB) outriggers, the Canadian Hydrogen Observatory and Radio-transient Detector (CHORD)\footnote{\url{https://www.chord-observatory.ca/home}}, and the Deep Synoptic Array (DSA)-2000\footnote{\url{https://www.deepsynoptic.org/}} are expected to detect $\sim 10^4$ localized FRBs per year (e.g., \cite{Vanderlinde2019,Leung2021,Lanman2024}). 
With the many FRBs available, the lensing bias must be carefully considered in the DM--$z$ relation. 

There are two types of FRBs: repeaters and non-repeaters.
Multiple signals from a repeater are consolidated into a single event when calculating $\langle {\rm DM} \rangle$ to prevent duplication.
In contrast, multiple images of a single source resulting from strong lensing are treated as independent events because their light ray paths differ.
However, the probability of such lensing occurrences is relatively low, e.g., $\approx 10^{-3}$ for lensing by a galaxy and $\approx 10^{-5}$ for lensing by a galaxy cluster (e.g., \cite{Oguri2019}).  

The numerical results in Fig.~\ref{fig_averageDM} can be fitted by a simple function:
\begin{equation}
    \frac{\langle {\rm DM} \rangle(z_{\rm s})}{\overline{\rm DM}(z_{\rm s})} = 1- \frac{1}{a z_{\rm s}^\gamma+b},
\end{equation}
where the second term corresponds to $\Delta {\rm DM}/\overline{\rm DM}$.
The fitting parameters are $(a,b,\gamma)=(37,63,-1.2)$ for TNG and $(a,b,\gamma)=(24,136,-1.6)$ for \textsc{HMx} with $\log_{10}(T_{\rm AGN}/{\rm K})=7.8$.
The relative difference between the numerical results and the fitting functions is less than $2.2 \times 10^{-4}$ and $3.8 \times 10^{-4}$ for TNG and \textsc{HMx}, respectively, over the range of $z_{\rm s}=0$--$3$.

\section{Magnification bias}
\label{sec:mag_bias}

Gravitational lensing in an inhomogeneous universe magnifies (or demagnifies) the flux of a source by a magnification factor $\mu$.
As a light ray traverses a higher (or lower) density region, both the DM and magnification increase (or decrease); thus, the DM correlates with magnification.
Because sources fainter than the detection threshold cannot be observed, more demagnified sources are likely to be missed compared to magnified ones. 
Therefore, the observed sample includes brighter, magnified events with higher DMs (or excludes fainter, demagnified events with lower DMs). 
This observational selection bias is known as magnification bias (e.g., \cite{Turner1984,SEF1992,BS2001}).
This section addresses this bias in $\langle {\rm DM} \rangle$.

The observed (lensed) luminosity $L_{\rm obs}$ and the unlensed luminosity $L$ are related as $L_{\rm obs} = \mu L$. 
Because the number density of sources is conserved with and without lensing magnification, we have $\Phi_{\rm obs}(L_{\rm obs};z_{\rm s}) {\rm d}L_{\rm obs} = \Phi(L;z_{\rm s}) {\rm d}L$, where $\Phi_{\rm obs} (\Phi)$ is the luminosity function with (without) lensing.
The lensed luminosity function is then expressed as
\begin{equation}
    \Phi_{\rm obs}(L_{\rm obs};z_{\rm s}) = 
    \int_{\mu_{\rm min}}^\infty \!\!
    \frac{{\rm d} \mu}{\mu} P_\mu(\mu;z_{\rm s}) \, \Phi \left( \frac{L_{\rm obs}}{\mu};z_{\rm s} \right),
\label{eq:lensed_LF}
\end{equation}
where $P_\mu$ is the probability distribution function of $\mu$, satisfying $\int {\rm d}\mu \, P_\mu=1$ and $\int {\rm d}\mu \, \mu \, P_\mu= \langle \mu \rangle = 1$.
The minimum magnification $\mu_{\rm min}$ is given as $\mu_{\rm min}=(1-\kappa_{\rm min})^{-2}$ where the minimum convergence $\kappa_{\rm min}$ is obtained in Eq.~(\ref{eq:kappa}) with $\delta_{\rm m}=-1$ (the empty beam).
The ensemble average of the DM with the magnification bias is obtained from Eq.~(\ref{eq:lensed_LF}):
\begin{equation}
\langle {\rm DM} \rangle_{\rm mag}(z_{\rm s}) = \frac{\displaystyle
\int_{\mu_{\rm min}}^\infty \!\! {\rm d} \mu \, \mu^{-1} \int_{L_{\rm obs,thr}}^\infty \!\!\!\!\! {\rm d}L_{\rm obs} \int_0^\infty \!\! {\rm dDM} \, P_{\mu,  {\rm DM}}(\mu,{\rm DM};z_{\rm s}) \, {\rm DM} \, \Phi(L_{\rm obs} \, \mu^{-1};z_{\rm s})}{\displaystyle  \int_{\mu_{\rm min}}^\infty \!\! {\rm d} \mu \, \mu^{-1} \int_{L_{\rm obs,thr}}^\infty \!\!\!\!\! {\rm d}L_{\rm obs} \, P_\mu(\mu;z_{\rm s}) \, \Phi(L_{\rm obs} \, \mu^{-1};z_{\rm s})},
\label{eq:ensemble_DM_mag}
\end{equation}
where $P_{\mu, {\rm DM}}$ is the joint probability distribution function of $\mu$ and the DM. 
If there is no correlation between them (i.e., $P_{\mu, {\rm DM}}(\mu,{\rm DM}) = P_\mu(\mu) \times P_{\rm DM}({\rm DM})$), 
$\langle {\rm DM} \rangle_{\rm mag}$ reduces to $\langle {\rm DM} \rangle$ in Eq.~(\ref{eq:ensemble_DM}).
The threshold luminosity $L_{\rm obs,thr}$ is determined by the flux limit $f_{\rm thr}$ as $L_{\rm obs,thr}=4 \pi D_{\rm L}^2 f_{\rm thr}$.
To calculate $\langle {\rm DM} \rangle_{\rm mag}$, a theoretical model of $P_{\mu,{\rm DM}}$ and $\Phi$ is necessary; however, $P_{\mu,{\rm DM}}$ has not yet been studied extensively (cosmological ray-tracing simulations would be required to obtain $P_{\mu,{\rm DM}}$).
Instead, we will calculate $\langle {\rm DM} \rangle_{\rm mag}$ under a few assumptions in the following.

Considering the luminosity function of a power-law form, $\Phi \propto L^{-\alpha}$ ($\alpha$ is a constant), $\langle {\rm DM} \rangle_{\rm mag}$ in Eq.~(\ref{eq:ensemble_DM_mag}) reduces to a simpler form:
\begin{align}
  \langle {\rm DM} \rangle_{\rm mag}(z_{\rm s}) &= \frac{\displaystyle \int_{\mu_{\rm min}}^\infty \!\! {\rm d} \mu  \int_0^\infty \!\! {\rm dDM} \, P_{\mu,  {\rm DM}}(\mu,{\rm DM};z_{\rm s}) \, {\mu}^{\alpha-1} \, {\rm DM}}{\displaystyle \int_{\mu_{\rm min}}^\infty \!\! {\rm d} \mu \, P_\mu(\mu;z_{\rm s}) \, \mu^{\alpha-1}}, \nonumber \\
  &= \frac{\langle \mu^{\alpha-1} {\rm DM} \rangle}{\langle \mu^{\alpha-1} \rangle}.
\label{eq:ensemble_DM_mag2}
\end{align}
This equation indicates that the magnification bias disappears when $\alpha=1$. 
We further use the weak lensing approximation, in which the magnification can be expanded as $\mu=[(1-\kappa)^2-\gamma^2]^{-1} \simeq 1+2\kappa+3\kappa^2+\gamma^2$ with $|\kappa|,|\gamma| \ll 1$.
Using $\langle \kappa^2 \rangle = \langle \gamma^2 \rangle$ and $\langle \kappa \rangle = -2 \langle \kappa^2 \rangle$, the denominator of Eq.~(\ref{eq:ensemble_DM_mag2}) is 
\begin{equation}
    \langle \mu^{\alpha-1} \rangle \simeq 1+ 2 \left( \alpha-1 \right) \left( \alpha-2 \right) \langle \kappa^2 \rangle.
    \nonumber
\end{equation}
Using Eqs.~(\ref{eq:ensemble_DM}) and (\ref{eq:deltaDM_me}), the numerator is 
\begin{equation}
    \langle \mu^{\alpha-1} {\rm DM} \rangle \simeq \overline{\rm DM} + 2 \left( \alpha-1 \right) \left( \alpha-2 \right) \langle \kappa^2 \rangle \, \overline{\rm DM} + \left( 2-\alpha \right) \Delta {\rm DM}.
    \nonumber
\end{equation}
Here, we have neglected the third and higher order terms of the perturbative quantities ($\kappa$, $\gamma$, and $\Delta {\rm DM}$).  
Finally, $\langle {\rm DM} \rangle_{\rm mag}$ simplifies to
\begin{equation}
  \langle {\rm DM} \rangle_{\rm mag}(z_{\rm s}) 
  \simeq \overline{\rm DM}(z_{\rm s}) + \Delta {\rm DM}(z_{\rm s}) +  \left( 1-\alpha \right) \Delta {\rm DM}(z_{\rm s}).
\label{eq:ensemble_DM_mag3}
\end{equation}
The second term represents the primary correction discussed in Section \ref{sec:2}, while the third accounts for magnification bias.
Both terms introduce a comparable bias ($\Delta {\rm DM}$).
For larger values of $\alpha$ ($>1$), there are more faint sub-threshold sources than super-threshold ones.    
Some of these sub-threshold sources become detectable owing to magnification, thereby increasing $\langle {\rm DM} \rangle_{\rm mag}$ (where $\Delta {\rm DM}<0$).
For $\alpha=2$, this increase cancels out the lensing bias in the second term of Eq.~(\ref{eq:ensemble_DM_mag3}).
The opposite is true for smaller values of $\alpha$ ($< 1$).
Some of the many super-threshold sources are lost owing to demagnification; thus, the magnification bias decreases $\langle {\rm DM} \rangle_{\rm mag}$.
A plot of $\langle {\rm DM} \rangle_{\rm mag}/\overline{\rm DM}$ as a function of $z_{\rm s}$ is the same as in Fig.~\ref{fig_averageDM}, but the deviation from unity is multiplied by a factor of $(2-\alpha)$. 

The index $\alpha$ of the FRB luminosity (or energy) function remains poorly constrained by current observations. 
Previous studies have predominantly explored either a single power-law function (e.g., \cite{Hashimoto2020,James2022b}) or a Schechter function\footnote{A Schechter function is the same as the power-law model but incorporating an exponential cutoff at a high $L$.} (e.g., \cite{Luo2020,Hashimoto2022,Shin2023}) to fit observational data, yielding best-fit values of $\alpha$ ranging broadly from $0.4$ to $1.8$. 
A more precise determination of $\alpha$ is needed to estimate $\langle {\rm DM} \rangle_{\rm mag}$.

\section{Discussion}
\label{sec:discussion}

\subsection{The other second-order effects on $\langle {\rm DM} \rangle$}
\label{subsec:discussion1}

We have examined the lensing bias affecting $\langle {\rm DM} \rangle$. 
However, other second-order terms in perturbation theory also affect $\langle {\rm DM} \rangle$.
According to previous studies concerning $\langle D_{\rm L} \rangle$ (e.g., \cite{Ben-Dayan2012,Ben-Dayan2013JCAP,Ben-Dayan2013PRL,Umeh2014}), the lensing bias dominates over other second-order effects for $z_{\rm s} \gtrsim 0.3$, suggesting a similar dominance for $\langle {\rm DM} \rangle$.
These additional factors include the Doppler effect from the peculiar motions of the observer and source, the integrated Sachs--Wolfe (ISW) effect owing to potential evolution along the photon path, and the SW effect arising from potential differences between the observer and source.
The peculiar velocity is approximately $v/c \approx 10^{-3} \, [v/(300 \, {\rm km/s})]$, and the galactic potential is $|\phi|/c^2 \approx 10^{-5}$--$10^{-6}$, both of which are smaller than the lensing convergence $|\kappa| \approx \langle \kappa^2 \rangle^{1/2} \approx 10^{-2} z_{\rm s}$.  
These effects introduce a slight bias on the source redshift of the order of $(v/c)^2$ for the peculiar velocity, $(\phi/c^2)^2$ for the potential term, and  $(v \phi /c^4)$ for their cross term (e.g., \cite{Umeh2014}). 
The quantitative assessment of all second-order effects on $\langle {\rm DM} \rangle$ is beyond the scope of this paper and remains a topic for future investigation.

\subsection{Rotation measure}
\label{subsec:discussion2}
The lensing bias can also be directly applied to the cosmological rotation measure (RM) (e.g., a review \cite{Han2017}).
When a radio signal propagates through a magnetized plasma, its polarization angle undergoes Faraday rotation (e.g., \cite{BinneyMerrifield1998}). 
The RM is derived from the frequency-dependent rotation angle.
Similar to the DM, the RM is expressed as (e.g., \cite{Akahori2011})
\begin{equation}
 {\rm RM}(z_{\rm s}) = \frac{e^3}{2 \pi m_{\rm e}^2 c^4} \int_0^{\lambda_{\rm s}} \! {\rm d}\lambda \, n_{\rm e}(\bm{r};z) \, B_\parallel(\bm{r};z) \, (1+z)^2,
\end{equation}
where $e$ and $m_{\rm e}$ are the electric charge and mass of the electron, respectively, and $B_\parallel$ is the comoving magnetic field component along the line of sight. 
Because light rays propagate through underdense regions owing to lensing, we expect that the lensing bias also influences $\langle \left| {\rm RM} \right| \rangle$\footnote{Here, $\langle {\rm RM} \rangle=0$ because $\langle B_\parallel \rangle=0$.}. 
Estimating the lensing bias in $\langle \left| {\rm RM} \right| \rangle$ requires the cross-power spectrum of matter and $n_{\rm e} B_\parallel$ fluctuations, which is beyond the scope of this study.

\section{Conclusions}
\label{sec:conclusion}

We investigated the lensing effects on the ensemble average of the cosmological dispersion measure, $\langle {\rm DM} \rangle$. 
Using the cosmological perturbation theory, we derived the primary correction term in $\langle {\rm DM} \rangle$ involving the cross-power spectrum of matter and free electrons. 
We found that $\langle {\rm DM} \rangle$ is $0.4 \, \%$--$1 \, \%$ smaller than the DM in the homogeneous cosmological model at $z_{\rm s}=1$, based on recent hydrodynamic simulations of IllustrisTNG and BAHAMAS.
The impact of lensing bias becomes more pronounced for higher $z_{\rm s}$ and in scenarios with weaker baryonic feedback models. 
Additionally, we discussed magnification bias, where brighter magnified sources with higher DM values are preferentially included in the observed sample. 
The magnification bias introduces a similar level of bias, $\mathcal{O}(0.1 \%$--$1 \%$), on $\langle {\rm DM} \rangle$ depending on the luminosity (or energy) function of FRBs.

Upcoming detectors, such as the CHIME/FRB outriggers, CHORD, and DSA-2000, are expected to detect $\approx 10^4$ localized FRBs annually.
During this era, it will be crucial to account for lensing bias, including magnification bias, to achieve accuracy at the percent level in the DM--$z$ relation.

\bibliographystyle{JHEP}
\bibliography{refs}

\providecommand{\href}[2]{#2}\begingroup\raggedright\begin{thebibliography}{100}

\bibitem{Bhandari2021}
S.~{Bhandari} and C.~{Flynn}, \emph{{Probing the Universe with Fast Radio
  Bursts}}, \href{https://doi.org/10.3390/universe7040085}{\emph{Universe}
  {\bfseries 7} (2021) 85}.

\bibitem{Xiao2021}
D.~{Xiao}, F.~{Wang} and Z.~{Dai}, \emph{{The physics of fast radio bursts}},
  \href{https://doi.org/10.1007/s11433-020-1661-7}{\emph{Science China Physics,
  Mechanics, and Astronomy} {\bfseries 64} (2021) 249501}
  [\href{https://arxiv.org/abs/arXiv:2101.04907}{{\ttfamily
  arXiv:2101.04907}}].

\bibitem{Petroff2022}
E.~{Petroff}, J.W.T.~{Hessels} and D.R.~{Lorimer}, \emph{{Fast radio bursts at
  the dawn of the 2020s}},
  \href{https://doi.org/10.1007/s00159-022-00139-w}{\emph{\aapr} {\bfseries 30}
  (2022) 2} [\href{https://arxiv.org/abs/arXiv:2107.10113}{{\ttfamily
  arXiv:2107.10113}}].

\bibitem{Zhang2023}
B.~{Zhang}, \emph{{The physics of fast radio bursts}},
  \href{https://doi.org/10.1103/RevModPhys.95.035005}{\emph{Reviews of Modern
  Physics} {\bfseries 95} (2023) 035005}
  [\href{https://arxiv.org/abs/arXiv:2212.03972}{{\ttfamily
  arXiv:2212.03972}}].

\bibitem{Keane2016}
E.F.~{Keane}, S.~{Johnston}, S.~{Bhandari}, E.~{Barr}, N.D.R.~{Bhat},
  M.~{Burgay} et~al., \emph{{The host galaxy of a fast radio burst}},
  \href{https://doi.org/10.1038/nature17140}{\emph{nature} {\bfseries 530}
  (2016) 453} [\href{https://arxiv.org/abs/arXiv:1602.07477}{{\ttfamily
  arXiv:1602.07477}}].

\bibitem{Chatterjee2017}
S.~{Chatterjee}, C.J.~{Law}, R.S.~{Wharton}, S.~{Burke-Spolaor},
  J.W.T.~{Hessels}, G.C.~{Bower} et~al., \emph{{A direct localization of a fast
  radio burst and its host}},
  \href{https://doi.org/10.1038/nature20797}{\emph{nature} {\bfseries 541}
  (2017) 58} [\href{https://arxiv.org/abs/arXiv:1701.01098}{{\ttfamily
  arXiv:1701.01098}}].

\bibitem{Bhandari2022}
S.~{Bhandari}, K.E.~{Heintz}, K.~{Aggarwal}, L.~{Marnoch}, C.K.~{Day},
  J.~{Sydnor} et~al., \emph{{Characterizing the Fast Radio Burst Host Galaxy
  Population and its Connection to Transients in the Local and Extragalactic
  Universe}}, \href{https://doi.org/10.3847/1538-3881/ac3aec}{\emph{\aj}
  {\bfseries 163} (2022) 69}
  [\href{https://arxiv.org/abs/arXiv:2108.01282}{{\ttfamily
  arXiv:2108.01282}}].

\bibitem{Bhardwaj2023}
M.~{Bhardwaj}, D.~{Michilli}, A.Y.~{Kirichenko}, O.~{Modilim}, K.~{Shin},
  V.M.~{Kaspi} et~al., \emph{{Host Galaxies for Four Nearby CHIME/FRB Sources
  and the Local Universe FRB Host Galaxy Population}},
  \href{https://doi.org/10.48550/arXiv.2310.10018}{\emph{arXiv e-prints} (2023)
  arXiv:2310.10018}.

\bibitem{Gordon2023}
A.C.~{Gordon}, W.-f.~{Fong}, C.D.~{Kilpatrick}, T.~{Eftekhari}, J.~{Leja},
  J.X.~{Prochaska} et~al., \emph{{The Demographics, Stellar Populations, and
  Star Formation Histories of Fast Radio Burst Host Galaxies: Implications for
  the Progenitors}},
  \href{https://doi.org/10.3847/1538-4357/ace5aa}{\emph{\apj} {\bfseries 954}
  (2023) 80} [\href{https://arxiv.org/abs/arXiv:2302.05465}{{\ttfamily
  arXiv:2302.05465}}].

\bibitem{Ibik2024}
A.L.~{Ibik}, M.R.~{Drout}, B.M.~{Gaensler}, P.~{Scholz}, D.~{Michilli},
  M.~{Bhardwaj} et~al., \emph{{Proposed Host Galaxies of Repeating Fast Radio
  Burst Sources Detected by CHIME/FRB}},
  \href{https://doi.org/10.3847/1538-4357/ad0893}{\emph{\apj} {\bfseries 961}
  (2024) 99} [\href{https://arxiv.org/abs/arXiv:2304.02638}{{\ttfamily
  arXiv:2304.02638}}].

\bibitem{Law2024}
C.J.~{Law}, K.~{Sharma}, V.~{Ravi}, G.~{Chen}, M.~{Catha}, L.~{Connor} et~al.,
  \emph{{Deep Synoptic Array Science: First FRB and Host Galaxy Catalog}},
  \href{https://doi.org/10.3847/1538-4357/ad3736}{\emph{\apj} {\bfseries 967}
  (2024) 29} [\href{https://arxiv.org/abs/arXiv:2307.03344}{{\ttfamily
  arXiv:2307.03344}}].

\bibitem{Ryder2023}
S.D.~{Ryder}, K.W.~{Bannister}, S.~{Bhandari}, A.T.~{Deller}, R.D.~{Ekers},
  M.~{Glowacki} et~al., \emph{{A luminous fast radio burst that probes the
  Universe at redshift 1}},
  \href{https://doi.org/10.1126/science.adf2678}{\emph{Science} {\bfseries 382}
  (2023) 294} [\href{https://arxiv.org/abs/arXiv:2210.04680}{{\ttfamily
  arXiv:2210.04680}}].

\bibitem{Li2020}
Z.~{Li}, H.~{Gao}, J.J.~{Wei}, Y.P.~{Yang}, B.~{Zhang} and Z.H.~{Zhu},
  \emph{{Cosmology-insensitive estimate of IGM baryon mass fraction from five
  localized fast radio bursts}},
  \href{https://doi.org/10.1093/mnrasl/slaa070}{\emph{\mnras} {\bfseries 496}
  (2020) L28} [\href{https://arxiv.org/abs/arXiv:2004.08393}{{\ttfamily
  arXiv:2004.08393}}].

\bibitem{Lemos2023}
T.~{Lemos}, R.~{Gon{\c{c}}alves}, J.~{Carvalho} and J.~{Alcaniz},
  \emph{{Forecasting constraints on the baryon mass fraction in the IGM from
  fast radio bursts and type Ia supernovae}},
  \href{https://doi.org/10.1140/epjc/s10052-023-12248-6}{\emph{European
  Physical Journal C} {\bfseries 83} (2023) 1128}
  [\href{https://arxiv.org/abs/arXiv:2307.06911}{{\ttfamily
  arXiv:2307.06911}}].

\bibitem{Lin2023}
H.-N.~{Lin} and R.~{Zou}, \emph{{Probing the baryon mass fraction in IGM and
  its redshift evolution with fast radio bursts using Bayesian inference
  method}}, \href{https://doi.org/10.1093/mnras/stad509}{\emph{\mnras}
  {\bfseries 520} (2023) 6237}
  [\href{https://arxiv.org/abs/arXiv:2302.10585}{{\ttfamily
  arXiv:2302.10585}}].

\bibitem{Wang2023}
B.~{Wang} and J.-J.~{Wei}, \emph{{An 8.0\% Determination of the Baryon Fraction
  in the Intergalactic Medium from Localized Fast Radio Bursts}},
  \href{https://doi.org/10.3847/1538-4357/acb2c8}{\emph{\apj} {\bfseries 944}
  (2023) 50} [\href{https://arxiv.org/abs/arXiv:2211.02209}{{\ttfamily
  arXiv:2211.02209}}].

\bibitem{Khrykin2024}
I.S.~{Khrykin}, M.~{Ata}, K.-G.~{Lee}, S.~{Simha}, Y.~{Huang}, J.X.~{Prochaska}
  et~al., \emph{{FLIMFLAM DR1: The First Constraints on the Cosmic Baryon
  Distribution from 8 FRB sightlines}},
  \href{https://doi.org/10.48550/arXiv.2402.00505}{\emph{arXiv e-prints} (2024)
  arXiv:2402.00505}.

\bibitem{Hagstotz2022}
S.~{Hagstotz}, R.~{Reischke} and R.~{Lilow}, \emph{{A new measurement of the
  Hubble constant using fast radio bursts}},
  \href{https://doi.org/10.1093/mnras/stac077}{\emph{\mnras} {\bfseries 511}
  (2022) 662} [\href{https://arxiv.org/abs/arXiv:2104.04538}{{\ttfamily
  arXiv:2104.04538}}].

\bibitem{James2022}
C.W.~{James}, E.M.~{Ghosh}, J.X.~{Prochaska}, K.W.~{Bannister}, S.~{Bhandari},
  C.K.~{Day} et~al., \emph{{A measurement of Hubble's Constant using Fast Radio
  Bursts}}, \href{https://doi.org/10.1093/mnras/stac2524}{\emph{\mnras}
  {\bfseries 516} (2022) 4862}
  [\href{https://arxiv.org/abs/arXiv:2208.00819}{{\ttfamily
  arXiv:2208.00819}}].

\bibitem{Wu2022}
Q.~{Wu}, G.-Q.~{Zhang} and F.-Y.~{Wang}, \emph{{An 8 per cent determination of
  the Hubble constant from localized fast radio bursts}},
  \href{https://doi.org/10.1093/mnrasl/slac022}{\emph{\mnras} {\bfseries 515}
  (2022) L1} [\href{https://arxiv.org/abs/arXiv:2108.00581}{{\ttfamily
  arXiv:2108.00581}}].

\bibitem{Fortunato2023}
J.A.S.~{Fortunato}, W.S.~{Hip{\'o}lito-Ricaldi} and M.V.~{dos Santos},
  \emph{{Cosmography from well-localized fast radio bursts}},
  \href{https://doi.org/10.1093/mnras/stad2856}{\emph{\mnras} {\bfseries 526}
  (2023) 1773} [\href{https://arxiv.org/abs/arXiv:2307.04711}{{\ttfamily
  arXiv:2307.04711}}].

\bibitem{Liu2023}
Y.~{Liu}, H.~{Yu} and P.~{Wu}, \emph{{Cosmological-model-independent
  Determination of Hubble Constant from Fast Radio Bursts and Hubble Parameter
  Measurements}}, \href{https://doi.org/10.3847/2041-8213/acc650}{\emph{\apjl}
  {\bfseries 946} (2023) L49}
  [\href{https://arxiv.org/abs/arXiv:2210.05202}{{\ttfamily
  arXiv:2210.05202}}].

\bibitem{Wei2023}
J.-J.~{Wei} and F.~{Melia}, \emph{{Investigating Cosmological Models and the
  Hubble Tension Using Localized Fast Radio Bursts}},
  \href{https://doi.org/10.3847/1538-4357/acefb8}{\emph{\apj} {\bfseries 955}
  (2023) 101} [\href{https://arxiv.org/abs/arXiv:2308.05918}{{\ttfamily
  arXiv:2308.05918}}].

\bibitem{Gao2024}
J.~{Gao}, Z.~{Zhou}, M.~{Du}, R.~{Zou}, J.~{Hu} and L.~{Xu}, \emph{{A
  measurement of Hubble constant using cosmographic approach combining fast
  radio bursts and supernovae}},
  \href{https://doi.org/10.1093/mnras/stad3708}{\emph{\mnras} {\bfseries 527}
  (2024) 7861} [\href{https://arxiv.org/abs/arXiv:2307.08285}{{\ttfamily
  arXiv:2307.08285}}].

\bibitem{Macquart2020}
J.P.~{Macquart}, J.X.~{Prochaska}, M.~{McQuinn}, K.W.~{Bannister},
  S.~{Bhandari}, C.K.~{Day} et~al., \emph{{A census of baryons in the Universe
  from localized fast radio bursts}},
  \href{https://doi.org/10.1038/s41586-020-2300-2}{\emph{nature} {\bfseries
  581} (2020) 391} [\href{https://arxiv.org/abs/arXiv:2005.13161}{{\ttfamily
  arXiv:2005.13161}}].

\bibitem{Yang2022}
K.B.~{Yang}, Q.~{Wu} and F.Y.~{Wang}, \emph{{Finding the Missing Baryons in the
  Intergalactic Medium with Localized Fast Radio Bursts}},
  \href{https://doi.org/10.3847/2041-8213/aca145}{\emph{\apjl} {\bfseries 940}
  (2022) L29} [\href{https://arxiv.org/abs/arXiv:2211.04058}{{\ttfamily
  arXiv:2211.04058}}].

\bibitem{Gao2014}
H.~{Gao}, Z.~{Li} and B.~{Zhang}, \emph{{Fast Radio Burst/Gamma-Ray Burst
  Cosmography}}, \href{https://doi.org/10.1088/0004-637X/788/2/189}{\emph{\apj}
  {\bfseries 788} (2014) 189}
  [\href{https://arxiv.org/abs/arXiv:1402.2498}{{\ttfamily arXiv:1402.2498}}].

\bibitem{Zhou2014}
B.~{Zhou}, X.~{Li}, T.~{Wang}, Y.-Z.~{Fan} and D.-M.~{Wei}, \emph{{Fast radio
  bursts as a cosmic probe?}},
  \href{https://doi.org/10.1103/PhysRevD.89.107303}{\emph{\prd} {\bfseries 89}
  (2014) 107303} [\href{https://arxiv.org/abs/arXiv:1401.2927}{{\ttfamily
  arXiv:1401.2927}}].

\bibitem{Walters2018}
A.~{Walters}, A.~{Weltman}, B.M.~{Gaensler}, Y.-Z.~{Ma} and A.~{Witzemann},
  \emph{{Future Cosmological Constraints From Fast Radio Bursts}},
  \href{https://doi.org/10.3847/1538-4357/aaaf6b}{\emph{\apj} {\bfseries 856}
  (2018) 65} [\href{https://arxiv.org/abs/arXiv:1711.11277}{{\ttfamily
  arXiv:1711.11277}}].

\bibitem{Ioka2003}
K.~{Ioka}, \emph{{The Cosmic Dispersion Measure from Gamma-Ray Burst
  Afterglows: Probing the Reionization History and the Burst Environment}},
  \href{https://doi.org/10.1086/380598}{\emph{\apjl} {\bfseries 598} (2003)
  L79} [\href{https://arxiv.org/abs/astro-ph/0309200}{{\ttfamily
  astro-ph/0309200}}].

\bibitem{Inoue2004}
S.~{Inoue}, \emph{{Probing the cosmic reionization history and local
  environment of gamma-ray bursts through radio dispersion}},
  \href{https://doi.org/10.1111/j.1365-2966.2004.07359.x}{\emph{\mnras}
  {\bfseries 348} (2004) 999}
  [\href{https://arxiv.org/abs/astro-ph/0309364}{{\ttfamily
  astro-ph/0309364}}].

\bibitem{Zheng2014}
Z.~{Zheng}, E.O.~{Ofek}, S.R.~{Kulkarni}, J.D.~{Neill} and M.~{Juric},
  \emph{{Probing the Intergalactic Medium with Fast Radio Bursts}},
  \href{https://doi.org/10.1088/0004-637X/797/1/71}{\emph{\apj} {\bfseries 797}
  (2014) 71} [\href{https://arxiv.org/abs/arXiv:1409.3244}{{\ttfamily
  arXiv:1409.3244}}].

\bibitem{Caleb2019}
M.~{Caleb}, C.~{Flynn} and B.W.~{Stappers}, \emph{{Constraining the era of
  helium reionization using fast radio bursts}},
  \href{https://doi.org/10.1093/mnras/stz571}{\emph{\mnras} {\bfseries 485}
  (2019) 2281} [\href{https://arxiv.org/abs/arXiv:1902.06981}{{\ttfamily
  arXiv:1902.06981}}].

\bibitem{Beniamini2021}
P.~{Beniamini}, P.~{Kumar}, X.~{Ma} and E.~{Quataert}, \emph{{Exploring the
  epoch of hydrogen reionization using FRBs}},
  \href{https://doi.org/10.1093/mnras/stab309}{\emph{\mnras} {\bfseries 502}
  (2021) 5134} [\href{https://arxiv.org/abs/arXiv:2011.11643}{{\ttfamily
  arXiv:2011.11643}}].

\bibitem{Hashimoto2021}
T.~{Hashimoto}, T.~{Goto}, T.-Y.~{Lu}, A.Y.L.~{On}, D.J.D.~{Santos}, S.J.~{Kim}
  et~al., \emph{{Revealing the cosmic reionization history with fast radio
  bursts in the era of Square Kilometre Array}},
  \href{https://doi.org/10.1093/mnras/stab186}{\emph{\mnras} {\bfseries 502}
  (2021) 2346} [\href{https://arxiv.org/abs/arXiv:2101.08798}{{\ttfamily
  arXiv:2101.08798}}].

\bibitem{Adamek2019}
J.~{Adamek}, C.~{Clarkson}, L.~{Coates}, R.~{Durrer} and M.~{Kunz}, \emph{{Bias
  and scatter in the Hubble diagram from cosmological large-scale structure}},
  \href{https://doi.org/10.1103/PhysRevD.100.021301}{\emph{\prd} {\bfseries
  100} (2019) 021301} [\href{https://arxiv.org/abs/arXiv:1812.04336}{{\ttfamily
  arXiv:1812.04336}}].

\bibitem{Barausse2005}
E.~{Barausse}, S.~{Matarrese} and A.~{Riotto}, \emph{{Effect of inhomogeneities
  on the luminosity distance-redshift relation: Is dark energy necessary in a
  perturbed universe?}},
  \href{https://doi.org/10.1103/PhysRevD.71.063537}{\emph{\prd} {\bfseries 71}
  (2005) 063537} [\href{https://arxiv.org/abs/astro-ph/0501152}{{\ttfamily
  astro-ph/0501152}}].

\bibitem{Ben-Dayan2012}
I.~{Ben-Dayan}, G.~{Marozzi}, F.~{Nugier} and G.~{Veneziano}, \emph{{The
  second-order luminosity-redshift relation in a generic inhomogeneous
  cosmology}},
  \href{https://doi.org/10.1088/1475-7516/2012/11/045}{\emph{\jcap} {\bfseries
  2012} (2012) 045} [\href{https://arxiv.org/abs/arXiv:1209.4326}{{\ttfamily
  arXiv:1209.4326}}].

\bibitem{Ben-Dayan2013JCAP}
I.~{Ben-Dayan}, M.~{Gasperini}, G.~{Marozzi}, F.~{Nugier} and G.~{Veneziano},
  \emph{{Average and dispersion of the luminosity-redshift relation in the
  concordance model}},
  \href{https://doi.org/10.1088/1475-7516/2013/06/002}{\emph{\jcap} {\bfseries
  2013} (2013) 002} [\href{https://arxiv.org/abs/arXiv:1302.0740}{{\ttfamily
  arXiv:1302.0740}}].

\bibitem{Ben-Dayan2013PRL}
I.~{Ben-Dayan}, M.~{Gasperini}, G.~{Marozzi}, F.~{Nugier} and G.~{Veneziano},
  \emph{{Do Stochastic Inhomogeneities Affect Dark-Energy Precision
  Measurements?}},
  \href{https://doi.org/10.1103/PhysRevLett.110.021301}{\emph{\prl} {\bfseries
  110} (2013) 021301} [\href{https://arxiv.org/abs/arXiv:1207.1286}{{\ttfamily
  arXiv:1207.1286}}].

\bibitem{Bonvin2015b}
C.~{Bonvin}, C.~{Clarkson}, R.~{Durrer}, R.~{Maartens} and O.~{Umeh}, \emph{{Do
  we care about the distance to the CMB? Clarifying the impact of second-order
  lensing}}, \href{https://doi.org/10.1088/1475-7516/2015/06/050}{\emph{\jcap}
  {\bfseries 2015} (2015) 050}
  [\href{https://arxiv.org/abs/arXiv:1503.07831}{{\ttfamily
  arXiv:1503.07831}}].

\bibitem{Brenton2021}
M.-A.~{Breton} and P.~{Fleury}, \emph{{Theoretical and numerical perspectives
  on cosmic distance averages}},
  \href{https://doi.org/10.1051/0004-6361/202040140}{\emph{\aap} {\bfseries
  655} (2021) A54} [\href{https://arxiv.org/abs/arXiv:2012.07802}{{\ttfamily
  arXiv:2012.07802}}].

\bibitem{Fleury2017}
P.~{Fleury}, C.~{Clarkson} and R.~{Maartens}, \emph{{How does the cosmic
  large-scale structure bias the Hubble diagram?}},
  \href{https://doi.org/10.1088/1475-7516/2017/03/062}{\emph{\jcap} {\bfseries
  2017} (2017) 062} [\href{https://arxiv.org/abs/arXiv:1612.03726}{{\ttfamily
  arXiv:1612.03726}}].

\bibitem{KP2016}
N.~{Kaiser} and J.A.~{Peacock}, \emph{{On the bias of the distance-redshift
  relation from gravitational lensing}},
  \href{https://doi.org/10.1093/mnras/stv2585}{\emph{\mnras} {\bfseries 455}
  (2016) 4518} [\href{https://arxiv.org/abs/arXiv:1503.08506}{{\ttfamily
  arXiv:1503.08506}}].

\bibitem{Marozzi2015}
G.~{Marozzi}, \emph{{The luminosity distance-redshift relation up to second
  order in the Poisson gauge with anisotropic stress}},
  \href{https://doi.org/10.1088/0264-9381/32/4/045004}{\emph{Classical and
  Quantum Gravity} {\bfseries 32} (2015) 045004}
  [\href{https://arxiv.org/abs/arXiv:1406.1135}{{\ttfamily arXiv:1406.1135}}].

\bibitem{Umeh2014}
O.~{Umeh}, C.~{Clarkson} and R.~{Maartens}, \emph{{Nonlinear relativistic
  corrections to cosmological distances, redshift and gravitational lensing
  magnification: I. Key results}},
  \href{https://doi.org/10.1088/0264-9381/31/20/202001}{\emph{Classical and
  Quantum Gravity} {\bfseries 31} (2014) 202001}
  [\href{https://arxiv.org/abs/arXiv:1207.2109}{{\ttfamily arXiv:1207.2109}}].

\bibitem{Umeh2014II}
O.~{Umeh}, C.~{Clarkson} and R.~{Maartens}, \emph{{Nonlinear relativistic
  corrections to cosmological distances, redshift and gravitational lensing
  magnification: II. Derivation}},
  \href{https://doi.org/10.1088/0264-9381/31/20/205001}{\emph{Classical and
  Quantum Gravity} {\bfseries 31} (2014) 205001}
  [\href{https://arxiv.org/abs/arXiv:1402.1933}{{\ttfamily arXiv:1402.1933}}].

\bibitem{Helbig2020}
P.~{Helbig}, \emph{{Calculation of distances in cosmological models with
  small-scale inhomogeneities and their use in observational cosmology: a
  review}}, \href{https://doi.org/10.21105/astro.1912.12269}{\emph{The Open
  Journal of Astrophysics} {\bfseries 3} (2020) 1}
  [\href{https://arxiv.org/abs/arXiv:1912.12269}{{\ttfamily
  arXiv:1912.12269}}].

\bibitem{Clarkson2014}
C.~{Clarkson}, O.~{Umeh}, R.~{Maartens} and R.~{Durrer}, \emph{{What is the
  distance to the CMB?}},
  \href{https://doi.org/10.1088/1475-7516/2014/11/036}{\emph{\jcap} {\bfseries
  2014} (2014) 036} [\href{https://arxiv.org/abs/arXiv:1405.7860}{{\ttfamily
  arXiv:1405.7860}}].

\bibitem{Holz2005}
D.E.~{Holz} and E.V.~{Linder}, \emph{{Safety in Numbers: Gravitational Lensing
  Degradation of the Luminosity Distance-Redshift Relation}},
  \href{https://doi.org/10.1086/432085}{\emph{\apj} {\bfseries 631} (2005) 678}
  [\href{https://arxiv.org/abs/astro-ph/0412173}{{\ttfamily
  astro-ph/0412173}}].

\bibitem{Jonsson2010}
J.~{J{\"o}nsson}, M.~{Sullivan}, I.~{Hook}, S.~{Basa}, R.~{Carlberg},
  A.~{Conley} et~al., \emph{{Constraining dark matter halo properties using
  lensed Supernova Legacy Survey supernovae}},
  \href{https://doi.org/10.1111/j.1365-2966.2010.16467.x}{\emph{\mnras}
  {\bfseries 405} (2010) 535}
  [\href{https://arxiv.org/abs/arXiv:1002.1374}{{\ttfamily arXiv:1002.1374}}].

\bibitem{Shah2022}
P.~{Shah}, P.~{Lemos} and O.~{Lahav}, \emph{{Weak-lensing magnification of Type
  Ia supernovae from the Pantheon sample}},
  \href{https://doi.org/10.1093/mnras/stac1746}{\emph{\mnras} {\bfseries 515}
  (2022) 2305} [\href{https://arxiv.org/abs/arXiv:2203.09865}{{\ttfamily
  arXiv:2203.09865}}].

\bibitem{Shah2024}
P.~{Shah}, T.M.~{Davis}, D.~{Bacon}, D.~{Brout}, J.~{Frieman}, L.~{Galbany}
  et~al., \emph{{The Dark Energy Survey : Detection of weak lensing
  magnification of supernovae and constraints on dark matter haloes}},
  \href{https://doi.org/10.1093/mnras/stae1515}{\emph{\mnras} (2024) }
  [\href{https://arxiv.org/abs/arXiv:2406.05047}{{\ttfamily
  arXiv:2406.05047}}].

\bibitem{Shah2023}
P.~{Shah}, P.~{Lemos} and O.~{Lahav}, \emph{{The impact of weak lensing on Type
  Ia supernovae luminosity distances}},
  \href{https://doi.org/10.1093/mnrasl/slad008}{\emph{\mnras} {\bfseries 520}
  (2023) L68} [\href{https://arxiv.org/abs/arXiv:2210.10688}{{\ttfamily
  arXiv:2210.10688}}].

\bibitem{Weinberg1976}
S.~{Weinberg}, \emph{{Apparent luminosities in a locally inhomogeneous
  universe.}}, \href{https://doi.org/10.1086/182216}{\emph{\apjl} {\bfseries
  208} (1976) L1}.

\bibitem{RT2011}
R.~{Takahashi}, M.~{Oguri}, M.~{Sato} and T.~{Hamana}, \emph{{Probability
  Distribution Functions of Cosmological Lensing: Convergence, Shear, and
  Magnification}},
  \href{https://doi.org/10.1088/0004-637X/742/1/15}{\emph{\apj} {\bfseries 742}
  (2011) 15} [\href{https://arxiv.org/abs/arXiv:1106.3823}{{\ttfamily
  arXiv:1106.3823}}].

\bibitem{Mizuno2023}
M.~{Mizuno} and T.~{Suyama}, \emph{{Weak lensing of gravitational waves in wave
  optics: Beyond the Born approximation}},
  \href{https://doi.org/10.1103/PhysRevD.108.043511}{\emph{\prd} {\bfseries
  108} (2023) 043511} [\href{https://arxiv.org/abs/arXiv:2210.02062}{{\ttfamily
  arXiv:2210.02062}}].

\bibitem{Mizuno2024}
M.~{Mizuno}, T.~{Suyama} and R.~{Takahashi}, \emph{{New consistency relations
  between averages and variances of weakly lensed signals of gravitational
  waves}}, \href{https://doi.org/10.1103/PhysRevD.109.083505}{\emph{\prd}
  {\bfseries 109} (2024) 083505}
  [\href{https://arxiv.org/abs/arXiv:2309.04114}{{\ttfamily
  arXiv:2309.04114}}].

\bibitem{McQuinn2014}
M.~{McQuinn}, \emph{{Locating the ``Missing'' Baryons with Extragalactic
  Dispersion Measure Estimates}},
  \href{https://doi.org/10.1088/2041-8205/780/2/L33}{\emph{\apjl} {\bfseries
  780} (2014) L33} [\href{https://arxiv.org/abs/arXiv:1309.4451}{{\ttfamily
  arXiv:1309.4451}}].

\bibitem{Dolag2015}
K.~{Dolag}, B.M.~{Gaensler}, A.M.~{Beck} and M.C.~{Beck}, \emph{{Constraints on
  the distribution and energetics of fast radio bursts using cosmological
  hydrodynamic simulations}},
  \href{https://doi.org/10.1093/mnras/stv1190}{\emph{\mnras} {\bfseries 451}
  (2015) 4277} [\href{https://arxiv.org/abs/arXiv:1412.4829}{{\ttfamily
  arXiv:1412.4829}}].

\bibitem{Shirasaki2017}
M.~{Shirasaki}, K.~{Kashiyama} and N.~{Yoshida}, \emph{{Large-scale clustering
  as a probe of the origin and the host environment of fast radio bursts}},
  \href{https://doi.org/10.1103/PhysRevD.95.083012}{\emph{\prd} {\bfseries 95}
  (2017) 083012} [\href{https://arxiv.org/abs/arXiv:1702.07085}{{\ttfamily
  arXiv:1702.07085}}].

\bibitem{Jaroszynski2019}
M.~{Jaroszynski}, \emph{{Fast radio bursts and cosmological tests}},
  \href{https://doi.org/10.1093/mnras/sty3529}{\emph{\mnras} {\bfseries 484}
  (2019) 1637} [\href{https://arxiv.org/abs/arXiv:1812.11936}{{\ttfamily
  arXiv:1812.11936}}].

\bibitem{TI2021}
R.~{Takahashi}, K.~{Ioka}, A.~{Mori} and K.~{Funahashi}, \emph{{Statistical
  modelling of the cosmological dispersion measure}},
  \href{https://doi.org/10.1093/mnras/stab170}{\emph{\mnras} {\bfseries 502}
  (2021) 2615} [\href{https://arxiv.org/abs/arXiv:2010.01560}{{\ttfamily
  arXiv:2010.01560}}].

\bibitem{Planck2016}
{Planck Collaboration}, \emph{{Planck 2015 results. XIII. Cosmological
  parameters}}, \href{https://doi.org/10.1051/0004-6361/201525830}{\emph{\aap}
  {\bfseries 594} (2016) A13}
  [\href{https://arxiv.org/abs/arXiv:1502.01589}{{\ttfamily
  arXiv:1502.01589}}].

\bibitem{BS2001}
M.~{Bartelmann} and P.~{Schneider}, \emph{{Weak gravitational lensing}},
  \href{https://doi.org/10.1016/S0370-1573(00)00082-X}{\emph{\physrep}
  {\bfseries 340} (2001) 291}
  [\href{https://arxiv.org/abs/astro-ph/9912508}{{\ttfamily
  astro-ph/9912508}}].

\bibitem{Dodelson2020}
S.~{Dodelson} and F.~{Schmidt}, \emph{{Modern Cosmology}} (2020),
  \href{https://doi.org/10.1016/C2017-0-01943-2}{10.1016/C2017-0-01943-2}.

\bibitem{Er2014}
X.~{Er} and S.~{Mao}, \emph{{Effects of plasma on gravitational lensing}},
  \href{https://doi.org/10.1093/mnras/stt2043}{\emph{\mnras} {\bfseries 437}
  (2014) 2180} [\href{https://arxiv.org/abs/arXiv:1310.5825}{{\ttfamily
  arXiv:1310.5825}}].

\bibitem{HS1965}
F.T.~{Haddock} and D.W.~{Sciama}, \emph{{Proposal for the Detection of
  Dispersion in Radio-Wave Propagation Through Intergalactic Space}},
  \href{https://doi.org/10.1103/PhysRevLett.14.1007}{\emph{\prl} {\bfseries 14}
  (1965) 1007}.

\bibitem{Ginzburg1973}
V.L.~{Ginzburg}, \emph{{Possibility of Determining Intergalactic Gas Density by
  Radio Observations of Flares of Remote Sources}},
  \href{https://doi.org/10.1038/246415a0}{\emph{nature} {\bfseries 246} (1973)
  415}.

\bibitem{Palmer1993}
D.M.~{Palmer}, \emph{{Radio Dispersion as a Diagnostic of Gamma-Ray Burst
  Distances}}, \href{https://doi.org/10.1086/187085}{\emph{\apjl} {\bfseries
  417} (1993) L25}.

\bibitem{DZ2014}
W.~{Deng} and B.~{Zhang}, \emph{{Cosmological Implications of Fast Radio
  Burst/Gamma-Ray Burst Associations}},
  \href{https://doi.org/10.1088/2041-8205/783/2/L35}{\emph{\apjl} {\bfseries
  783} (2014) L35} [\href{https://arxiv.org/abs/arXiv:1401.0059}{{\ttfamily
  arXiv:1401.0059}}].

\bibitem{Mead2020}
A.J.~{Mead}, T.~{Tr{\"o}ster}, C.~{Heymans}, L.~{Van Waerbeke} and
  I.G.~{McCarthy}, \emph{{A hydrodynamical halo model for weak-lensing cross
  correlations}},
  \href{https://doi.org/10.1051/0004-6361/202038308}{\emph{\aap} {\bfseries
  641} (2020) A130} [\href{https://arxiv.org/abs/arXiv:2005.00009}{{\ttfamily
  arXiv:2005.00009}}].

\bibitem{Marinancci2018}
F.~{Marinacci}, M.~{Vogelsberger}, R.~{Pakmor}, P.~{Torrey}, V.~{Springel},
  L.~{Hernquist} et~al., \emph{{First results from the IllustrisTNG
  simulations: radio haloes and magnetic fields}},
  \href{https://doi.org/10.1093/mnras/sty2206}{\emph{\mnras} {\bfseries 480}
  (2018) 5113} [\href{https://arxiv.org/abs/arXiv:1707.03396}{{\ttfamily
  arXiv:1707.03396}}].

\bibitem{Naiman2018}
J.P.~{Naiman}, A.~{Pillepich}, V.~{Springel}, E.~{Ramirez-Ruiz}, P.~{Torrey},
  M.~{Vogelsberger} et~al., \emph{{First results from the IllustrisTNG
  simulations: a tale of two elements - chemical evolution of magnesium and
  europium}}, \href{https://doi.org/10.1093/mnras/sty618}{\emph{\mnras}
  {\bfseries 477} (2018) 1206}
  [\href{https://arxiv.org/abs/arXiv:1707.03401}{{\ttfamily
  arXiv:1707.03401}}].

\bibitem{Nelson2018}
D.~{Nelson}, A.~{Pillepich}, V.~{Springel}, R.~{Weinberger}, L.~{Hernquist},
  R.~{Pakmor} et~al., \emph{{First results from the IllustrisTNG simulations:
  the galaxy colour bimodality}},
  \href{https://doi.org/10.1093/mnras/stx3040}{\emph{\mnras} {\bfseries 475}
  (2018) 624} [\href{https://arxiv.org/abs/arXiv:1707.03395}{{\ttfamily
  arXiv:1707.03395}}].

\bibitem{Phillepich2018}
A.~{Pillepich}, D.~{Nelson}, L.~{Hernquist}, V.~{Springel}, R.~{Pakmor},
  P.~{Torrey} et~al., \emph{{First results from the IllustrisTNG simulations:
  the stellar mass content of groups and clusters of galaxies}},
  \href{https://doi.org/10.1093/mnras/stx3112}{\emph{\mnras} {\bfseries 475}
  (2018) 648} [\href{https://arxiv.org/abs/arXiv:1707.03406}{{\ttfamily
  arXiv:1707.03406}}].

\bibitem{Springel2018}
V.~{Springel}, R.~{Pakmor}, A.~{Pillepich}, R.~{Weinberger}, D.~{Nelson},
  L.~{Hernquist} et~al., \emph{{First results from the IllustrisTNG
  simulations: matter and galaxy clustering}},
  \href{https://doi.org/10.1093/mnras/stx3304}{\emph{\mnras} {\bfseries 475}
  (2018) 676} [\href{https://arxiv.org/abs/arXiv:1707.03397}{{\ttfamily
  arXiv:1707.03397}}].

\bibitem{Nelson2019}
D.~{Nelson}, V.~{Springel}, A.~{Pillepich}, V.~{Rodriguez-Gomez}, P.~{Torrey},
  S.~{Genel} et~al., \emph{{The IllustrisTNG simulations: public data
  release}},
  \href{https://doi.org/10.1186/s40668-019-0028-x}{\emph{Computational
  Astrophysics and Cosmology} {\bfseries 6} (2019) 2}
  [\href{https://arxiv.org/abs/arXiv:1812.05609}{{\ttfamily
  arXiv:1812.05609}}].

\bibitem{Troster2022}
T.~{Tr{\"o}ster}, A.J.~{Mead}, C.~{Heymans}, Z.~{Yan}, D.~{Alonso}, M.~{Asgari}
  et~al., \emph{{Joint constraints on cosmology and the impact of baryon
  feedback: Combining KiDS-1000 lensing with the thermal Sunyaev-Zeldovich
  effect from Planck and ACT}},
  \href{https://doi.org/10.1051/0004-6361/202142197}{\emph{\aap} {\bfseries
  660} (2022) A27} [\href{https://arxiv.org/abs/arXiv:2109.04458}{{\ttfamily
  arXiv:2109.04458}}].

\bibitem{McCarthy2017}
I.G.~{McCarthy}, J.~{Schaye}, S.~{Bird} and A.M.C.~{Le Brun}, \emph{{The
  BAHAMAS project: calibrated hydrodynamical simulations for large-scale
  structure cosmology}},
  \href{https://doi.org/10.1093/mnras/stw2792}{\emph{\mnras} {\bfseries 465}
  (2017) 2936} [\href{https://arxiv.org/abs/arXiv:1603.02702}{{\ttfamily
  arXiv:1603.02702}}].

\bibitem{McMarthy2018}
I.G.~{McCarthy}, S.~{Bird}, J.~{Schaye}, J.~{Harnois-Deraps}, A.S.~{Font} and
  L.~{van Waerbeke}, \emph{{The BAHAMAS project: the CMB-large-scale structure
  tension and the roles of massive neutrinos and galaxy formation}},
  \href{https://doi.org/10.1093/mnras/sty377}{\emph{\mnras} {\bfseries 476}
  (2018) 2999} [\href{https://arxiv.org/abs/arXiv:1712.02411}{{\ttfamily
  arXiv:1712.02411}}].

\bibitem{Springel2010}
V.~{Springel}, \emph{{E pur si muove: Galilean-invariant cosmological
  hydrodynamical simulations on a moving mesh}},
  \href{https://doi.org/10.1111/j.1365-2966.2009.15715.x}{\emph{\mnras}
  {\bfseries 401} (2010) 791}
  [\href{https://arxiv.org/abs/arXiv:0901.4107}{{\ttfamily arXiv:0901.4107}}].

\bibitem{HE1981}
R.W.~{Hockney} and J.W.~{Eastwood}, \emph{{Computer Simulation Using Particles
  (New York: McGraw-Hill)}} (1981).

\bibitem{Jing2005}
Y.P.~{Jing}, \emph{{Correcting for the Alias Effect When Measuring the Power
  Spectrum Using a Fast Fourier Transform}},
  \href{https://doi.org/10.1086/427087}{\emph{\apj} {\bfseries 620} (2005) 559}
  [\href{https://arxiv.org/abs/astro-ph/0409240}{{\ttfamily
  astro-ph/0409240}}].

\bibitem{Sefusatti2016}
E.~{Sefusatti}, M.~{Crocce}, R.~{Scoccimarro} and H.M.P.~{Couchman},
  \emph{{Accurate estimators of correlation functions in Fourier space}},
  \href{https://doi.org/10.1093/mnras/stw1229}{\emph{\mnras} (2016) }
  [\href{https://arxiv.org/abs/arXiv:1512.07295}{{\ttfamily
  arXiv:1512.07295}}].

\bibitem{Springel2005}
V.~{Springel}, \emph{{The cosmological simulation code GADGET-2}},
  \href{https://doi.org/10.1111/j.1365-2966.2005.09655.x}{\emph{\mnras}
  {\bfseries 364} (2005) 1105}
  [\href{https://arxiv.org/abs/astro-ph/0505010}{{\ttfamily
  astro-ph/0505010}}].

\bibitem{CS2002}
A.~{Cooray} and R.~{Sheth}, \emph{{Halo models of large scale structure}},
  \href{https://doi.org/10.1016/S0370-1573(02)00276-4}{\emph{\physrep}
  {\bfseries 372} (2002) 1}
  [\href{https://arxiv.org/abs/astro-ph/0206508}{{\ttfamily
  astro-ph/0206508}}].

\bibitem{Arico2020}
G.~{Aric{\`o}}, R.E.~{Angulo}, C.~{Hern{\'a}ndez-Monteagudo}, S.~{Contreras},
  M.~{Zennaro}, M.~{Pellejero-Iba{\~n}ez} et~al., \emph{{Modelling the
  large-scale mass density field of the universe as a function of cosmology and
  baryonic physics}},
  \href{https://doi.org/10.1093/mnras/staa1478}{\emph{\mnras} {\bfseries 495}
  (2020) 4800} [\href{https://arxiv.org/abs/arXiv:1911.08471}{{\ttfamily
  arXiv:1911.08471}}].

\bibitem{Shirasaki2022}
M.~{Shirasaki}, R.~{Takahashi}, K.~{Osato} and K.~{Ioka}, \emph{{Probing
  cosmology and gastrophysics with fast radio bursts: cross-correlations of
  dark matter haloes and cosmic dispersion measures}},
  \href{https://doi.org/10.1093/mnras/stac490}{\emph{\mnras} {\bfseries 512}
  (2022) 1730} [\href{https://arxiv.org/abs/arXiv:2108.12205}{{\ttfamily
  arXiv:2108.12205}}].

\bibitem{Asgari2023}
M.~{Asgari}, A.J.~{Mead} and C.~{Heymans}, \emph{{The halo model for cosmology:
  a pedagogical review}},
  \href{https://doi.org/10.21105/astro.2303.08752}{\emph{The Open Journal of
  Astrophysics} {\bfseries 6} (2023) 39}
  [\href{https://arxiv.org/abs/arXiv:2303.08752}{{\ttfamily
  arXiv:2303.08752}}].

\bibitem{Reischke2023}
R.~{Reischke} and S.~{Hagstotz}, \emph{{Cosmological covariance of fast radio
  burst dispersions}},
  \href{https://doi.org/10.1093/mnras/stad1645}{\emph{\mnras} {\bfseries 524}
  (2023) 2237} [\href{https://arxiv.org/abs/arXiv:2301.03527}{{\ttfamily
  arXiv:2301.03527}}].

\bibitem{Reischke2023b}
R.~{Reischke}, D.~{Neumann}, K.A.~{Bertmann}, S.~{Hagstotz} and
  H.~{Hildebrandt}, \emph{{Calibrating baryonic feedback with weak lensing and
  fast radio bursts}},
  \href{https://doi.org/10.48550/arXiv.2309.09766}{\emph{arXiv e-prints} (2023)
  arXiv:2309.09766}.

\bibitem{Smith2003}
R.E.~{Smith}, J.A.~{Peacock}, A.~{Jenkins}, S.D.M.~{White}, C.S.~{Frenk},
  F.R.~{Pearce} et~al., \emph{{Stable clustering, the halo model and non-linear
  cosmological power spectra}},
  \href{https://doi.org/10.1046/j.1365-8711.2003.06503.x}{\emph{\mnras}
  {\bfseries 341} (2003) 1311}
  [\href{https://arxiv.org/abs/astro-ph/0207664}{{\ttfamily
  astro-ph/0207664}}].

\bibitem{RT2012}
R.~{Takahashi}, M.~{Sato}, T.~{Nishimichi}, A.~{Taruya} and M.~{Oguri},
  \emph{{Revising the Halofit Model for the Nonlinear Matter Power Spectrum}},
  \href{https://doi.org/10.1088/0004-637X/761/2/152}{\emph{\apj} {\bfseries
  761} (2012) 152} [\href{https://arxiv.org/abs/arXiv:1208.2701}{{\ttfamily
  arXiv:1208.2701}}].

\bibitem{Chisari2019}
N.E.~{Chisari}, A.J.~{Mead}, S.~{Joudaki}, P.G.~{Ferreira}, A.~{Schneider},
  J.~{Mohr} et~al., \emph{{Modelling baryonic feedback for survey cosmology}},
  \href{https://doi.org/10.21105/astro.1905.06082}{\emph{The Open Journal of
  Astrophysics} {\bfseries 2} (2019) 4}
  [\href{https://arxiv.org/abs/arXiv:1905.06082}{{\ttfamily
  arXiv:1905.06082}}].

\bibitem{Chen2023}
A.~{Chen}, G.~{Aric{\`o}}, D.~{Huterer}, R.E.~{Angulo}, N.~{Weaverdyck},
  O.~{Friedrich} et~al., \emph{{Constraining the baryonic feedback with cosmic
  shear using the DES Year-3 small-scale measurements}},
  \href{https://doi.org/10.1093/mnras/stac3213}{\emph{\mnras} {\bfseries 518}
  (2023) 5340} [\href{https://arxiv.org/abs/arXiv:2206.08591}{{\ttfamily
  arXiv:2206.08591}}].

\bibitem{Arico2023}
G.~{Aric{\`o}}, R.E.~{Angulo}, M.~{Zennaro}, S.~{Contreras}, A.~{Chen} and
  C.~{Hern{\'a}ndez-Monteagudo}, \emph{{DES Y3 cosmic shear down to small
  scales: Constraints on cosmology and baryons}},
  \href{https://doi.org/10.1051/0004-6361/202346539}{\emph{\aap} {\bfseries
  678} (2023) A109} [\href{https://arxiv.org/abs/arXiv:2303.05537}{{\ttfamily
  arXiv:2303.05537}}].

\bibitem{Xu2023}
J.~{Xu}, T.~{Eifler}, V.~{Miranda}, X.~{Fang}, E.~{Saraivanov}, E.~{Krause}
  et~al., \emph{{Constraining Baryonic Physics with DES Y1 and Planck data --
  Combining Galaxy Clustering, Weak Lensing, and CMB Lensing}}, {\emph{arXiv
  e-prints} (2023) arXiv:2311.08047}
  [\href{https://arxiv.org/abs/arXiv:2311.08047}{{\ttfamily
  arXiv:2311.08047}}].

\bibitem{Terasawa2024}
R.~{Terasawa}, X.~{Li}, M.~{Takada}, T.~{Nishimichi}, S.~{Tanaka},
  S.~{Sugiyama} et~al., \emph{{Exploring the baryonic effect signature in the
  Hyper Suprime-Cam Year 3 cosmic shear two-point correlations on small scales:
  the $S_8$ tension remains present}},
  \href{https://doi.org/10.48550/arXiv.2403.20323}{\emph{arXiv e-prints} (2024)
  arXiv:2403.20323} [\href{https://arxiv.org/abs/2403.20323}{{\ttfamily
  2403.20323}}].

\bibitem{Garcia2024}
C.~{Garc{\'\i}a-Garc{\'\i}a}, M.~{Zennaro}, G.~{Aric{\`o}}, D.~{Alonso} and
  R.E.~{Angulo}, \emph{{Cosmic shear with small scales: DES-Y3, KiDS-1000 and
  HSC-DR1}}, \href{https://doi.org/10.48550/arXiv.2403.13794}{\emph{arXiv
  e-prints} (2024) arXiv:2403.13794}
  [\href{https://arxiv.org/abs/2403.13794}{{\ttfamily 2403.13794}}].

\bibitem{Ferreira2023}
T.~{Ferreira}, D.~{Alonso}, C.~{Garcia-Garcia} and N.E.~{Chisari},
  \emph{{X-Ray-Cosmic-Shear Cross-Correlations: First Detection and Constraints
  on Baryonic Effects}},
  \href{https://doi.org/10.1103/PhysRevLett.133.051001}{\emph{\prl} {\bfseries
  133} (2024) 051001} [\href{https://arxiv.org/abs/arXiv:2309.11129}{{\ttfamily
  arXiv:2309.11129}}].

\bibitem{Kumar2019}
P.~{Kumar} and E.V.~{Linder}, \emph{{Use of fast radio burst dispersion
  measures as distance measures}},
  \href{https://doi.org/10.1103/PhysRevD.100.083533}{\emph{\prd} {\bfseries
  100} (2019) 083533} [\href{https://arxiv.org/abs/arXiv:1903.08175}{{\ttfamily
  arXiv:1903.08175}}].

\bibitem{Mo2023}
J.-F.~{Mo}, W.~{Zhu}, Y.~{Wang}, L.~{Tang} and L.-L.~{Feng}, \emph{{The
  dispersion measure of Fast Radio Bursts host galaxies: estimation from
  cosmological simulations}},
  \href{https://doi.org/10.1093/mnras/stac3104}{\emph{\mnras} {\bfseries 518}
  (2023) 539} [\href{https://arxiv.org/abs/arXiv:2210.14052}{{\ttfamily
  arXiv:2210.14052}}].

\bibitem{Theis2024}
A.~{Theis}, S.~{Hagstotz}, R.~{Reischke} and J.~{Weller}, \emph{{Galaxy
  dispersion measured by Fast Radio Bursts as a probe of baryonic feedback
  models}}, \href{https://doi.org/10.48550/arXiv.2403.08611}{\emph{arXiv
  e-prints} (2024) arXiv:2403.08611}.

\bibitem{Vanderlinde2019}
K.~{Vanderlinde}, A.~{Liu}, B.~{Gaensler}, D.~{Bond}, G.~{Hinshaw}, C.~{Ng}
  et~al., \emph{{The Canadian Hydrogen Observatory and Radio-transient Detector
  (CHORD)}},  in \emph{Canadian Long Range Plan for Astronomy and Astrophysics
  White Papers}, vol.~2020, p.~28, 2019,
  \href{https://doi.org/10.5281/zenodo.3765414}{DOI}
  [\href{https://arxiv.org/abs/arXiv:1911.01777}{{\ttfamily
  arXiv:1911.01777}}].

\bibitem{Leung2021}
C.~{Leung}, J.~{Mena-Parra}, K.~{Masui}, K.~{Bandura}, M.~{Bhardwaj},
  P.J.~{Boyle} et~al., \emph{{A Synoptic VLBI Technique for Localizing
  Nonrepeating Fast Radio Bursts with CHIME/FRB}},
  \href{https://doi.org/10.3847/1538-3881/abd174}{\emph{\aj} {\bfseries 161}
  (2021) 81} [\href{https://arxiv.org/abs/arXiv:2008.11738}{{\ttfamily
  arXiv:2008.11738}}].

\bibitem{Lanman2024}
A.E.~{Lanman}, S.~{Andrew}, M.~{Lazda}, V.~{Shah}, M.~{Amiri},
  A.~{Balasubramanian} et~al., \emph{{CHIME/FRB Outriggers: KKO Station System
  and Commissioning Results}},
  \href{https://doi.org/10.48550/arXiv.2402.07898}{\emph{arXiv e-prints} (2024)
  arXiv:2402.07898}.

\bibitem{Oguri2019}
M.~{Oguri}, \emph{{Strong gravitational lensing of explosive transients}},
  \href{https://doi.org/10.1088/1361-6633/ab4fc5}{\emph{Reports on Progress in
  Physics} {\bfseries 82} (2019) 126901}
  [\href{https://arxiv.org/abs/arXiv:1907.06830}{{\ttfamily
  arXiv:1907.06830}}].

\bibitem{Turner1984}
E.L.~{Turner}, J.P.~{Ostriker} and I.~{Gott}, J.~R., \emph{{The statistics of
  gravitational lenses : the distributions of image angular separations and
  lens redshifts.}}, \href{https://doi.org/10.1086/162379}{\emph{\apj}
  {\bfseries 284} (1984) 1}.

\bibitem{SEF1992}
P.~{Schneider}, J.~{Ehlers} and E.E.~{Falco}, \emph{{Gravitational Lenses}}
  (1992),
  \href{https://doi.org/10.1007/978-3-662-03758-4}{10.1007/978-3-662-03758-4}.

\bibitem{Hashimoto2020}
T.~{Hashimoto}, T.~{Goto}, A.Y.L.~{On}, T.-Y.~{Lu}, D.J.D.~{Santos},
  S.C.C.~{Ho} et~al., \emph{{No redshift evolution of non-repeating fast radio
  burst rates}}, \href{https://doi.org/10.1093/mnras/staa2490}{\emph{\mnras}
  {\bfseries 498} (2020) 3927}
  [\href{https://arxiv.org/abs/arXiv:2008.09621}{{\ttfamily
  arXiv:2008.09621}}].

\bibitem{James2022b}
C.W.~{James}, J.X.~{Prochaska}, J.P.~{Macquart}, F.O.~{North-Hickey},
  K.W.~{Bannister} and A.~{Dunning}, \emph{{The z-DM distribution of fast radio
  bursts}}, \href{https://doi.org/10.1093/mnras/stab3051}{\emph{\mnras}
  {\bfseries 509} (2022) 4775}
  [\href{https://arxiv.org/abs/arXiv:2101.08005}{{\ttfamily
  arXiv:2101.08005}}].

\bibitem{Luo2020}
R.~{Luo}, Y.~{Men}, K.~{Lee}, W.~{Wang}, D.R.~{Lorimer} and B.~{Zhang},
  \emph{{On the FRB luminosity function - - II. Event rate density}},
  \href{https://doi.org/10.1093/mnras/staa704}{\emph{\mnras} {\bfseries 494}
  (2020) 665} [\href{https://arxiv.org/abs/arXiv:2003.04848}{{\ttfamily
  arXiv:2003.04848}}].

\bibitem{Hashimoto2022}
T.~{Hashimoto}, T.~{Goto}, B.H.~{Chen}, S.C.C.~{Ho}, T.Y.Y.~{Hsiao},
  Y.H.V.~{Wong} et~al., \emph{{Energy functions of fast radio bursts derived
  from the first CHIME/FRB catalogue}},
  \href{https://doi.org/10.1093/mnras/stac065}{\emph{\mnras} {\bfseries 511}
  (2022) 1961} [\href{https://arxiv.org/abs/arXiv:2201.03574}{{\ttfamily
  arXiv:2201.03574}}].

\bibitem{Shin2023}
K.~{Shin}, K.W.~{Masui}, M.~{Bhardwaj}, T.~{Cassanelli}, P.~{Chawla},
  M.~{Dobbs} et~al., \emph{{Inferring the Energy and Distance Distributions of
  Fast Radio Bursts Using the First CHIME/FRB Catalog}},
  \href{https://doi.org/10.3847/1538-4357/acaf06}{\emph{\apj} {\bfseries 944}
  (2023) 105} [\href{https://arxiv.org/abs/arXiv:2207.14316}{{\ttfamily
  arXiv:2207.14316}}].

\bibitem{Han2017}
J.L.~{Han}, \emph{{Observing Interstellar and Intergalactic Magnetic Fields}},
  \href{https://doi.org/10.1146/annurev-astro-091916-055221}{\emph{\araa}
  {\bfseries 55} (2017) 111}.

\bibitem{BinneyMerrifield1998}
J.~{Binney} and M.~{Merrifield}, \emph{{Galactic Astronomy}} (1998).

\bibitem{Akahori2011}
T.~{Akahori} and D.~{Ryu}, \emph{{Faraday Rotation Measure due to the
  Intergalactic Magnetic Field. II. The Cosmological Contribution}},
  \href{https://doi.org/10.1088/0004-637X/738/2/134}{\emph{\apj} {\bfseries
  738} (2011) 134} [\href{https://arxiv.org/abs/arXiv:1107.0142}{{\ttfamily
  arXiv:1107.0142}}].

\bibitem{Bonvin2015}
C.~{Bonvin}, C.~{Clarkson}, R.~{Durrer}, R.~{Maartens} and O.~{Umeh},
  \emph{{Cosmological ensemble and directional averages of observables}},
  \href{https://doi.org/10.1088/1475-7516/2015/07/040}{\emph{\jcap} {\bfseries
  2015} (2015) 040} [\href{https://arxiv.org/abs/arXiv:1504.01676}{{\ttfamily
  arXiv:1504.01676}}].

\end{thebibliography}\endgroup

\appendix

\section{Derivation of the ${\rm DM}$ correction terms}
\label{sec:appendix1}

This appendix provides a detailed calculation of the DM correction terms in Eq.~({\ref{eq:expand_DM}}).  
The second and third terms of Eq.~(\ref{eq:expand_DM}) are written as $\Delta {\rm DM}_{\rm 2nd} (z_{\rm s})$ and  $\Delta {\rm DM}_{\rm 3rd} (z_{\rm s})$, respectively.
Substituting $\bm{x}$ in Eq.~(\ref{eq:geodesic}) into Eq.~(\ref{eq:expand_DM}), the second term becomes
\begin{align}
    \Delta {\rm DM}_{\rm 2nd}(z_{\rm s})  & =
    \frac{2}{c^2} \int_0^{r_{\rm s}} \! {\rm d}r \, \bar{n}_{\rm e}(z) \, (1+z)  
    \left[ \, - \! \int_0^r \!\! {\rm d}r^\prime \, ( r-r^\prime ) + \frac{r}{r_{\rm s}} \int_0^{r_{\rm s}} \!\! {\rm d}r^\prime \, ( r_{\rm s}-r^\prime ) \right] \nonumber \\
    &~~~ \times \left. \left\langle \nabla_{\bm{x}^\prime} \phi(\bm{r}^\prime;z^\prime) \cdot \nabla_{\bm{x}} \delta_{\rm e}(\bm{r};z) \right\rangle \right|_{\bm{x}=\bm{x}^\prime=0}.
\label{eq:dDM2nd}
\end{align}
Here, the integration is performed along the $r$-axis, and $\lambda, \lambda^\prime$ and $\lambda_{\rm s}$ have been substituted with $r,r^\prime$ and $r_{\rm s}$, respectively.

The Fourier transforms of $\phi$ and $\delta_{\rm e}$ are
\begin{equation}
    \phi(\bm{r};z) = \int \! \frac{{\rm d}^3k}{(2 \pi)^3} \tilde{\phi}(\bm{k};z) \, {\rm e}^{{\rm i} \bm{k} \cdot \bm{r}}, ~~\delta_{\rm e}(\bm{r};z) = \int \! \frac{{\rm d}^3k}{(2 \pi)^3} \tilde{\delta}_{\rm e}(\bm{k};z) \, {\rm e}^{{\rm i} \bm{k} \cdot \bm{r}}.
\label{eq:FT}
\end{equation}
The gravitational potential $\phi$ is determined by the matter density fluctuations $\delta_{\rm m}$ through the Poisson equation in Fourier space: $-k^2 \tilde{\phi}(\bm{k};z) = (3/2) H_0^2 \Omega_{\rm m} (1+z) \tilde {\delta}_{\rm m}(\bm{k};z)$.   
The cross correlation of $\nabla_{\bm{x}} \phi$ and $\nabla_{\bm{x}} \delta_{\rm e}$ in Eq.~(\ref{eq:dDM2nd}) is rewritten using the Fourier transforms (\ref{eq:FT}):
\begin{align}
   \left. \left\langle \nabla_{\bm{x}^\prime} \phi(\bm{r}^\prime;z^\prime) \cdot \nabla_{\bm{x}} \delta_{\rm e}(\bm{r};z) \right\rangle \right|_{\bm{x}=\bm{x}^\prime=0} 
   =& -\frac{3 H_0^2 \Omega_{\rm m}}{2} (1+z) \int \! \frac{{\rm d}^3k}{(2 \pi)^3}  \int \! \frac{{\rm d}^3k^\prime}{(2 \pi)^3} \frac{\bm{k}_\perp \cdot \bm{k}^\prime_\perp}{{k^\prime}^2} \nonumber \\ 
   &~~ \times \left\langle \tilde{\delta}_{\rm m}(\bm{k}^\prime;z^\prime) \tilde{\delta}_{\rm e}^* (\bm{k};z) \right\rangle 
   {\rm e}^{{\rm i} (k_\parallel^\prime r^\prime- k_\parallel r)}, 
\label{eq:cross_phi-ne}
\end{align}
where $k_\parallel$ and $\bm{k}_\perp$ are the wave vectors parallel and perpendicular to the $r$-axis (i.e., $\bm{k} \cdot \bm{r} = k_\parallel r + \bm{k}_\perp \cdot \bm{x}$), respectively. 
The cross-power spectrum between $\delta_{\rm m}$ and $\delta_{\rm e}$ is defined as
\begin{equation}
    \left\langle \tilde{\delta}_{\rm m}(\bm{k}^\prime;z^\prime) \tilde{\delta}_{\rm e}^* (\bm{k};z) \right\rangle \equiv (2 \pi)^3 \delta_{\rm D} (\bm{k}-\bm{k}^\prime) P_{\rm me} (k;z,z^\prime),
\label{eq:cross_pk}
\end{equation}
where $\delta_{\rm D}$ is the Dirac delta function.
Using Eq.~(\ref{eq:cross_pk}), Eq.~(\ref{eq:cross_phi-ne}) reduces to
\begin{align}
      \left. \left\langle \nabla_{\bm{x}^\prime} \phi(\bm{r}^\prime;z^\prime) \cdot \nabla_{\bm{x}} \delta_{\rm e}(\bm{r};z) \right\rangle \right|_{\bm{x}=\bm{x}^\prime=0} 
     &= -\frac{3 H_0^2 \Omega_{\rm m}}{2} (1+z) \int \! \frac{{\rm d}^3k}{(2 \pi)^3} \frac{k_\perp^2}{k^2} P_{\rm me}(k;z,z^\prime) \, {\rm e}^{{\rm i} k_\parallel (r^\prime -r)}, \nonumber \\
     &= -\frac{3 H_0^2 \Omega_{\rm m}}{2} (1+z) \int \! \frac{{\rm d}^2k_\perp}{(2 \pi)^2} P_{\rm me}(k_\perp;z) \, \delta_{\rm D} (r^\prime -r).
\label{eq:cross_phi-ne2}
\end{align}
In the second equality, we applied the Limber approximation: $\int {\rm d}k_\parallel \, {\rm e}^{{\rm i} k_\parallel (r^\prime -r)} P_{\rm me}(k;z,z^\prime)  = 2 \pi \delta_{\rm D}(r^\prime-r) P_{\rm me}(k_\perp;z)$, which neglects the cross-correlation of $\phi(\bm{r}^\prime;z^\prime)$ and $\delta_{\rm e}(\bm{r};z)$ for the different redshifts $z^\prime \neq z$. 
Substituting Eq.~(\ref{eq:cross_phi-ne2}) into Eq.~(\ref{eq:dDM2nd}) yields the following: 
\begin{align}
    \Delta {\rm DM}_{\rm 2nd}(z_{\rm s})  
    &=  -\frac{3H_0^2 \Omega_{\rm m}}{2 \pi c^2} \int_0^{r_{\rm s}} \! {\rm d}r  \, \bar{n}_{\rm e}(z) \, (1+z)^2 \frac{r (r_{\rm s}-r)}{r_{\rm s}} 
    \int_0^\infty \! {\rm d}k_\perp k_\perp P_{\rm me}(k_\perp;z), \nonumber \\
    &= -2 \, \left\langle \kappa \left( {\rm DM} -\overline{\rm DM} \right) \right\rangle. 
\label{eq:deltaDM_me}
\end{align}
Here, the convergence field for the source redshift $z_{\rm s}$ is defined as
\begin{equation}
    \kappa(z_{\rm s}) = \frac{3 H_0^2 \Omega_{\rm m}}{2 c^2} \int_0^{r_{\rm s}} \!\! {\rm d}r \, (1+z) \frac{r (r_{\rm s}-r)}{r_{\rm s}} \, \delta_{\rm m}(\bm{r};z).
\label{eq:kappa}
\end{equation}
Therefore, the correction term $\Delta {\rm DM}_{\rm 2nd}$ arises from a cross-correlation between $\kappa$ and the DM. 
Equation (\ref{eq:deltaDM_me}) is consistent with the previous work (Section 2 of Ref.~\cite{Bonvin2015}) by replacing their $f$ with the DM. 
Equations (\ref{eq:deltaDM_me}) and (\ref{eq:ensemble_DM}) agree with the previous result for the convergence~\cite{RT2011,KP2016}, $\langle \kappa \rangle = -2 \langle \kappa^2 \rangle$, by replacing $\bar{n}_{\rm e} \delta_{\rm e}$ with $(3 H_0^2 \Omega_{\rm m}/2 c^2) \, [r (r_{\rm s}-r)/r_{\rm s}] \, \delta_{\rm m}$ in Eqs.~(\ref{eq:DM})--(\ref{eq:ne}).

Next, we proceed to the third term of Eq.~(\ref{eq:expand_DM}).
This term arises from the difference in path length from the observer to the source between the straight line (along the $r$-axis) and the curved path (specified by $\lambda$) in Fig.~\ref{fig_config}. 
Substituting ${\rm d} \bm{x}/{\rm d}\lambda$ in Eq.~(\ref{eq:geodesic}) into Eq.~(\ref{eq:expand_DM}), this term becomes
\begin{align}
    \Delta {\rm DM}_{\rm 3rd}(z_{\rm s})  
    &= \frac{2}{c^4} \int_0^{r_{\rm s}} \! {\rm d}r \, \bar{n}_{\rm e}(z) \, (1+z) 
    \left[ \int_0^r \!  {\rm d}r^\prime \! \int_0^r \!  {\rm d}r^{\prime \prime} \right. 
    +\frac{1}{r_{\rm s}^2} \int_0^{r_{\rm s}} \!  {\rm d}r^\prime  (r_{\rm s}-r^\prime) \int_0^{r_{\rm s}} \!  {\rm d}r^{\prime \prime} (r_{\rm s}-r^{\prime\prime}) \nonumber \\
    &~~ \left. -\frac{2}{r_{\rm s}} \int_0^r \!  {\rm d}r^\prime \! \int_0^{r_{\rm s}} \!  {\rm d}r^{\prime \prime} (r_{\rm s}-r^{\prime\prime}) \right]  
    \left. \left\langle \nabla_{\bm{x}^\prime} \phi(\bm{r}^\prime;z^\prime) \cdot \nabla_{\bm{x}^{\prime \prime}} \phi(\bm{r}^{\prime \prime};z^{\prime \prime}) \right\rangle \right|_{\bm{x}^{\prime}=\bm{x}^{\prime\prime}=0} .
\label{eq:dDM3rd}
\end{align}
Similar to Eq.~(\ref{eq:cross_phi-ne2}), the correlation function of $\nabla_{\bm{x}} \phi$ in Eq.~(\ref{eq:dDM3rd}) is obtained as follows:
\begin{align}
    & \left. \left\langle \nabla_{\bm{x}^\prime} \phi(\bm{r}^\prime;z^\prime) \cdot \nabla_{\bm{x}^{\prime \prime}} \phi(\bm{r}^{\prime \prime};z^{\prime \prime}) \right\rangle \right|_{\bm{x}^{\prime}=\bm{x}^{\prime\prime}=0} 
    = \frac{9 H_0^4 \Omega_{\rm m}^2}{4} (1+z^\prime)^2 \! \int \frac{{\rm d}^2k_\perp}{(2 \pi)^2} \, \frac{1}{k_\perp^2} P_{\rm m}(k_\perp;z^\prime) \, \delta_{\rm D}(r^\prime-r^{\prime\prime}),
\label{eq:cross_phi-phi}
\end{align}
where $P_{\rm m}$ is the matter power spectrum defined as 
    $\langle \tilde{\delta}_{\rm m}(\bm{k}^\prime;z) \tilde{\delta}_{\rm m}^* (\bm{k};z)\rangle \equiv (2 \pi)^3 \delta_{\rm D} (\bm{k}-\bm{k}^\prime) P_{\rm m} (k;z)$.
Using Eq.~(\ref{eq:cross_phi-phi}), Eq.~(\ref{eq:dDM3rd}) is rewritten as
\begin{align}
    \Delta {\rm DM}_{\rm 3rd}(z_{\rm s})  = \int_0^{r_{\rm s}} \! {\rm d}r \, \bar{n}_{\rm e}(z) \, (1+z) \int_0^r \! {\rm d}r^\prime \, \frac{2r^\prime - r_{\rm s}}{r_{\rm s}} \, J(z^\prime)  
    +\overline{\rm DM}(z_{\rm s}) \int_0^{r_{\rm s}} \! {\rm d}r \left( \frac{r_{\rm s}-r}{r_{\rm s}} \right)^2 \!\! J(z), 
\end{align}
with
\begin{equation}
    J(z) = \frac{9 H_0^4 \Omega_{\rm m}^2}{4 \pi c^4} (1+z)^2 \int_0^\infty \frac{{\rm d}k_\perp}{k_\perp} P_{\rm m}(k_\perp;z).
\end{equation}
The function $J(z)$ is the same as in KP16.
The smallness of the dimensionless quantity $r J$ is discussed in Section 1.3 of KP16.
The third term $\Delta {\rm DM}_{\rm 3rd}$ is orders of magnitude smaller compared to the first and second terms in Eq.~(\ref{eq:expand_DM}).

\section{Free-electron bias}
\label{sec:be}

\begin{figure}
  \centering
  \includegraphics[width=15.5cm, clip]{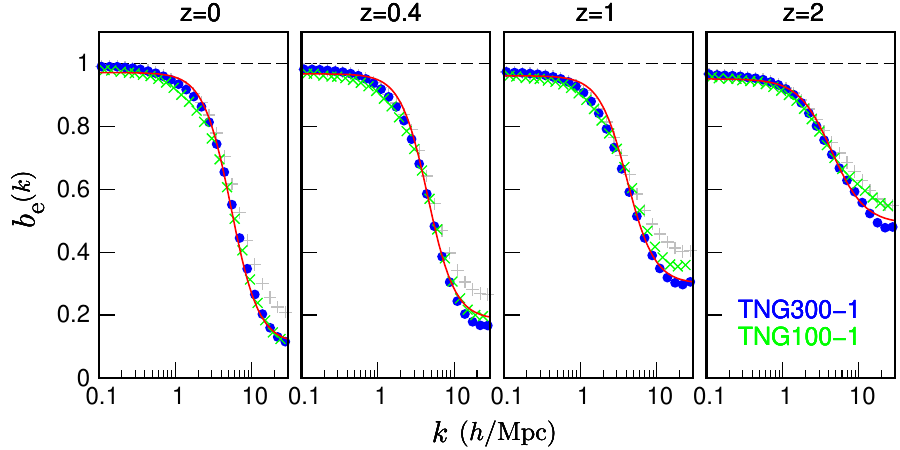}  
  \caption{
    The free-electron bias is defined as the ratio of the cross-power spectrum $P_{\rm me}(k)$ to the matter power spectrum in the DMO run $P_{\rm DMO}(k)$: $b_{\rm e}(k) \equiv P_{\rm me}(k)/P_{\rm DMO}(k)$.
    The blue and green symbols represent measurements in TNG300-1 and TNG100-1, respectively.
    The solid red curve is the fitting function (\ref{eq:be}). 
    The gray pluses are the bias in our previous work~\cite{TI2021}, $b_{\rm e}^{\rm (prev)}(k;z) \equiv [P_{\rm e}(k;z)/P_{\rm DMO}(k;z)]^{1/2}$, measured in TNG300-1. 
    }
  \label{fig_pk-ratio-DMO}
\end{figure}

\begin{figure}
  \centering
  \includegraphics[width=15.5cm, clip]{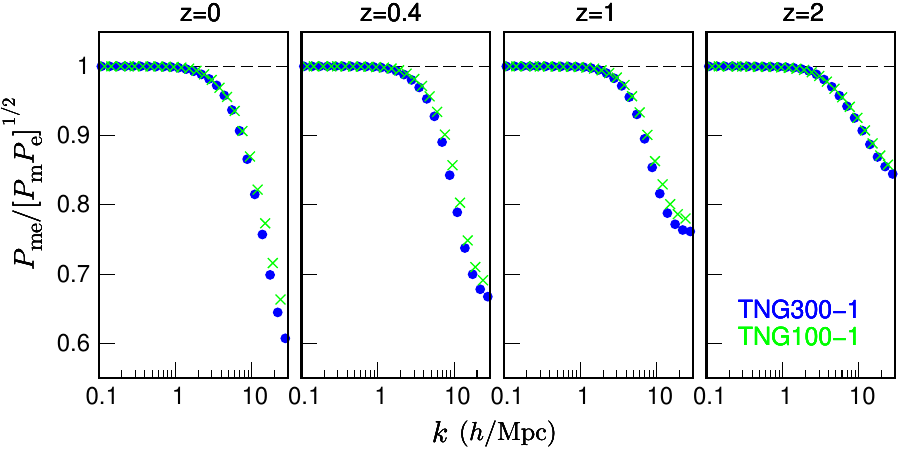}   
  \caption{
    The cross-correlation coefficient is defined as $P_{\rm me}(k)/[P_{\rm m}(k)P_{\rm e}(k)]^{1/2}$.
    The blue and green symbols represent measurements in TNG300-1 and TNG100-1, respectively.
    }
  \label{fig_pk-ccc}
\end{figure}

This section introduces a fitting function for the free-electron bias, defined as $b_{\rm e}(k;z)$ $\equiv P_{\rm me}(k;z)/P_{\rm DMO}(k;z)$, using TNG300-1 and TNG100-1.
The fitting range is $k<38.3 \, h \, {\rm Mpc}^{-1}$ and $76.2 \, h \, {\rm Mpc}^{-1}$ for TNG300-1 and TNG100-1 (i.e., the Nyquist wavenumbers), respectively, in $z=0$--$3$ ($z=0,0.2,0.4,0.7,1,2$ and $3$).
Figure \ref{fig_pk-ratio-DMO} shows the bias for $z=0,0.4,1$ and $2$.
The sample variance of the power spectrum, caused by the finite simulation volume, is canceled because the hydrodynamic run and the DMO run have the same initial seed of random Gaussian fluctuations.
The bias is fitted as follows:
\begin{equation}
    b_{\rm e}(k;z)=\frac{b_{0}(z)}{1+[ k/k_*(z)]^\gamma}+b_{1}(z),
\label{eq:be}
\end{equation}
with
\begin{align}
    b_0(z) &= 0.86-0.20 z,   \nonumber  \\
    b_1(z) &= 0.11+0.19 z,   \nonumber  \\
    k_*(z) &= 5.4-2.9 z+2.0 z^2-0.40 z^3, \nonumber \\
    \gamma(z) &= 2.4+0.2 z-0.2 z^2, \nonumber
\end{align}
where $k_*$ is in unit of $h \, {\rm Mpc}^{-1}$.
The fitting function agrees with both simulations (TNG300-1 and TNG100-1) to within $4.3 \, \%$ ($9.5 \, \%$) for $k \leq 1$ ($10$) $h \,{\rm Mpc}^{-1}$ and $z=0$--$3$. 
The bias approaches unity on large scales ($k \lesssim 1 \, h \, {\rm Mpc}^{-1}$), but is significantly reduced on small scales ($k \gtrsim 1 \, h \, {\rm Mpc}^{-1}$) owing to AGN and stellar feedback.
The bias is slightly smaller than unity on the largest scales, especially for higher redshifts, because the fluctuations in baryons gradually catch up to the fluctuations in dark matter after the cosmological recombination (e.g., Subsection 3.3 of Ref.~\cite{TI2021}).

It is worth noting that the free-electron bias differs slightly from the bias defined in our previous study~\cite{TI2021} using the free-electron power spectrum $P_{\rm e}$: $b_{\rm e}^{\rm (prev)}(k;z)=$ $[P_{\rm e}(k;z)$ $/P_{\rm DMO}(k;z)]^{1/2}$.
The gray pluses in Fig.~\ref{fig_pk-ratio-DMO} represent the previous bias.
Both biases are nearly identical on large scales ($k \lesssim 1 \, h \, {\rm Mpc}^{-1}$), but the new one is slightly more suppressed at $k \, \gtrsim \, 1 h {\rm Mpc}^{-1}$.    
To view this difference from another perspective, the cross-correlation coefficient,
\begin{equation}
    r(k;z) \equiv \frac{P_{\rm me}(k;z)}{\left[ P_{\rm m}(k;z) P_{\rm e}(k;z) \right]^{1/2}},
\end{equation}
is plotted in Fig.~\ref{fig_pk-ccc}.
Here, $P_{\rm m}$ is the matter power spectrum of the hydrodynamic simulations (not the DMO simulations).
The coefficient $r$ quantifies the phase difference between the two fields.
If the phases differ, such as when $\arg (\tilde{\delta}_{\rm e}(\bm{k};z)) - \arg (\tilde{\delta}_{\rm m}(\bm{k};z)) = \theta$ where $\theta$ is a constant, then $r=\cos \theta$.
Alternatively, if $\tilde{\delta}_{\rm e}(\bm{k};z) = b_{1}(k;z) \, \tilde{\delta}_{\rm m}(\bm{k};z)$ with a real function $b_{1}$, then $r=1$. 
Figure \ref{fig_pk-ccc} shows $r=1$ for large scales ($k \lesssim 1 \, h \, {\rm Mpc}^{-1}$), but $r<1$ for small scales ($k \gtrsim 1 \, h \, {\rm Mpc}^{-1}$). 
Here, $r<1$ leads to additional small-scale suppression of $b_{\rm e}$ compared to $b_{\rm e}^{\rm (prev)}$, as shown in Fig.~\ref{fig_pk-ratio-DMO}.

\acknowledgments

We thank Kunihito Ioka, Alex Mead, and Masamune Oguri for their helpful comments and suggestions. 
We thank the Enago (www.enago.jp) editors for the English language review.
This work is supported in part by JSPS KAKENHI grant Nos. JP22H00130 and JP20H05855.

\end{document}